# Extending Single Molecule Förster Resonance Energy Transfer (FRET) Range Beyond 10 Nanometers in Zero-Mode Waveguides

Mikhail Baibakov,[1,*] Satyajit Patra,[1,*] Jean-Benoît Claude,[1] Antonin Moreau,[1] Julien Lumeau,[1] Jérôme Wenger[1,#]

[1] Aix Marseille Univ, CNRS, Centrale Marseille, Institut Fresnel, 13013 Marseille, France

[*] These authors contributed equally to this work

[#] Corresponding author: jerome.wenger@fresnel.fr

**Abstract**

Single molecule Förster resonance energy transfer (smFRET) is widely used to monitor conformations and interactions dynamics at the molecular level. However, conventional smFRET measurements are ineffective at donor-acceptor distances exceeding 10 nm, impeding the studies on biomolecules of larger size. Here, we show that zero-mode waveguide (ZMW) apertures can be used to overcome the 10 nm barrier in smFRET. Using an optimized ZMW structure, we demonstrate smFRET between standard commercial fluorophores up to 13.6 nm distance with a significantly improved FRET efficiency. To further break into the classical FRET range limit, ZMWs are combined with molecular constructs featuring multiple acceptor dyes to achieve high FRET efficiencies together with high fluorescence count rates. As we discuss general guidelines for quantitative smFRET measurements inside ZMWs, the technique can be readily applied for monitoring conformations and interactions on large molecular complexes with enhanced brightness.

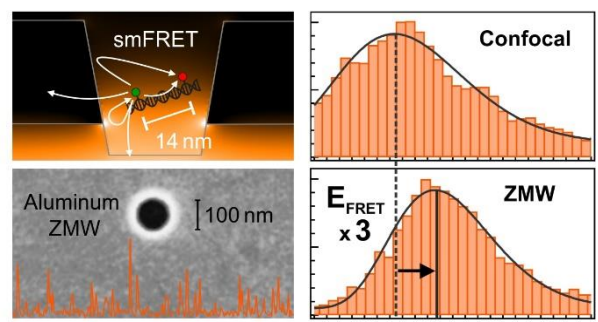

Figure for Table of Contents

Förster resonance energy transfer (FRET) accounts for the energy transfer between a donor and an acceptor fluorescent emitter, and is widely used to monitor biomolecular conformations and interactions dynamics.[1–3] As the energy transfer rate decays as the inverse 6th power of the donor-acceptor separation, single molecule FRET (smFRET) is highly sensitive to the relative distance between the fluorophores and has therefore been termed a molecular ruler, enabling accurate distance measurements on the 3-9 nm scale.[4] However, at donor-acceptor separations greater than 10 nm, the FRET efficiency falls below 10%. Detecting FRET becomes then highly challenging using conventional single molecule approaches: the lifetime-based FRET measurements fail to distinguish between donor-acceptor and donor-only labeled samples, while the intensity-based FRET measurements struggle with very low acceptor count rates, donor emission leaking into the acceptor detection channel and incomplete fluorescence labelling.[1,2]

New approaches are needed to fully exploit the potential of smFRET also for large biomolecular constructs beyond the 10 nm barrier.[5] So far, most attention has been focused on designing elaborate donor-acceptor constructs to extend the FRET range. This involves the use of long-lifetime lanthanides as donors,[6,7] multi-color cascaded FRET systems,[8,9] gold nanoparticles quenchers as acceptors,[10–12] or multiple donor or acceptor dyes to further promote the energy transfer.[13–16] However, all these approaches suffer from either low photon count rates, complex sample preparation and/or advanced instrumentation, which hinders their applicability to a broad set of DNAs and proteins.[17]

To overcome these limits, an alternative strategy uses nanophotonic components to tailor the electromagnetic environment surrounding the FRET pair in such a way to promote the energy transfer and enhance the fluorescence detection rates.[18–21] A significant advantage of this approach is that it preserves the ability to use standard FRET fluorophore pairs. Several contributions have explored the influence of nanophotonics for FRET using microcavities,[18,22–24] mirrors,[25–29] nanoparticles,[30–37] nanoapertures,[38–44] nanoantennas,[45–49] waveguides,[50] or hyperbolic metamaterials.[51,52] However, none of these works has clearly demonstrated experimentally the enhancement of the smFRET detection range in the near field. Actually, most cases consider short donor-acceptor separations (on the order or below the Förster radius) in order to ease the optical detection.[23,25,36,37,46,47,52] It should be acknowledged that long range energy transfer over distances up to several micrometers has been reported,[53–58] but these studies are based on radiative dipole-dipole coupling (*i.e.* energy transfer mediated by a propagating photon or plasmon in the far field). This situation is fundamentally different from FRET which involves near-field dipole-dipole coupling *via* evanescent waves. A striking difference between FRET and radiative coupling is that in the former case there is a change in the donor lifetime induced by the presence of the acceptor while in the latter case there is none.

In addition to the extension of the smFRET range, a second major challenge for nanophotonics is to improve the detected FRET efficiency. Although the nanophotonic structure enhances the FRET rate, it simultaneously also increases the other donor radiative and non-radiative processes, which are directly competing with FRET.[36,37,47,52] The net result is that the FRET efficiency is often significantly quenched by the presence of the nanophotonic structure.[23,25,37,46,47]

Here, we describe an optimized approach to detect smFRET between standard organic fluorophores over distances exceeding 10 nm with enhanced efficiency. An essential element is the zero-mode waveguide (ZMW,[59–62] a single nanoaperture of 100 nm diameter milled in an aluminum film, see Fig. 1a,b), which replaces the glass coverslip generally used in single molecule FRET microscopy and confines the light at a spatial scale of a few tens of nanometers. Thanks to the fluorescence enhancement occurring in the ZMW, single molecule FRET can be clearly detected at 13.6 nm separation between Atto550-Atto647N FRET pairs on double-stranded DNA samples, and with a significant 3-fold improvement on the detected FRET efficiency. This result is enabled by the combination of two key elements: (i) improved nanofabrication of aluminum nanostructures featuring lower losses and higher fluorescence enhancement factors and (ii) pulsed interleaved excitation (PIE-FRET) to clearly resolve the single FRET pairs, perform quantitative analysis and avoid the issues related to incomplete fluorescence labelling.[4,63] The earlier works from our group[39,40] and others[38,41–43] on FRET with ZMWs lacked these two key ingredients and therefore could not demonstrate clearly the enhancement of the FRET efficiency at long distances. With this significant step forward, we can now provide general guidelines to optimize the design of zero-mode waveguides for smFRET experiments overcoming the 10 nm barrier. We also investigate the combination of constructs featuring multiple identical acceptor dyes with ZMWs in order to further extend the FRET detection range while preserving a high FRET efficiency and high count rates. With their smooth circular shape, aluminum-based ZMWs are quite easy to fabricate with standard electron or ion lithography and feature an optical response covering the full visible spectral range. Therefore, the results shown here establish a broadly applicable method to extend the FRET detection range and improve the collection statistics on almost every fluorophore construct and confocal microscope.

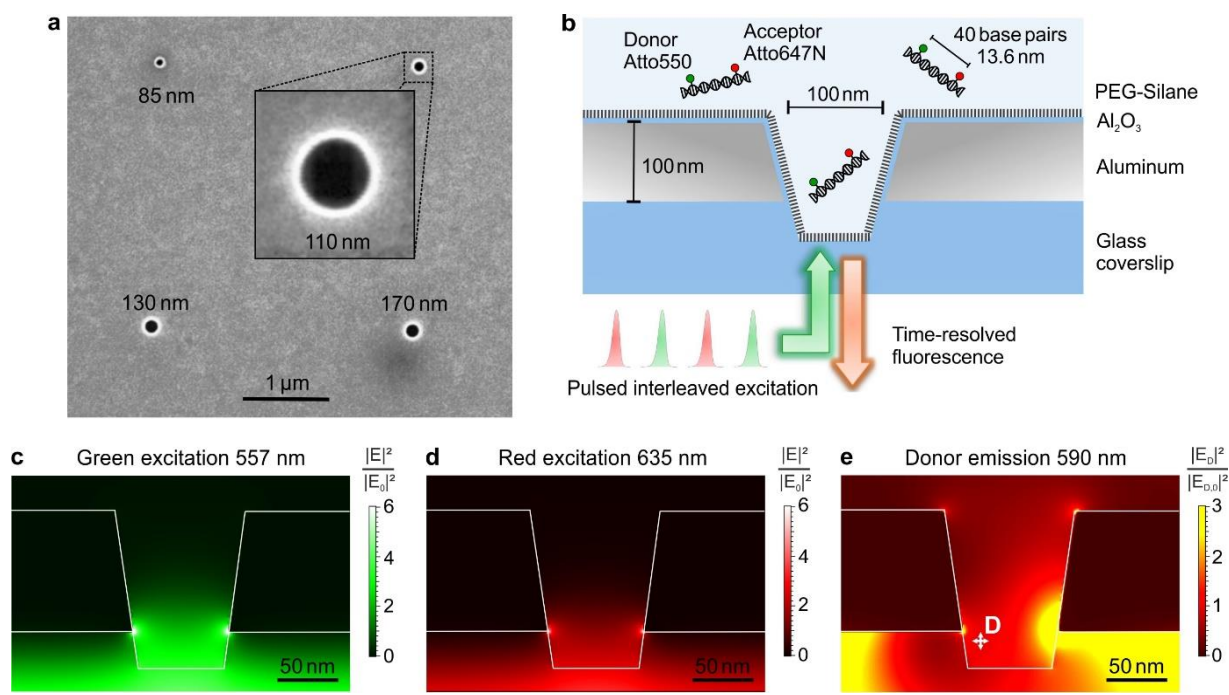

**Figure 1.** Pulsed interleaved excitation of single molecule FRET in zero-mode waveguides. (a) Scanning electron microscope image of zero-mode waveguides with different diameters used in this study. (b) Scheme of the experiment: two alternating pulsed laser beams are focused below a single zero-mode waveguide (ZMW) to sequentially detect fluorescence from FRET, donor direct excitation and acceptor direct excitation. Single FRET pairs on double stranded DNA are diffusing across the ZMW. (c-e) Numerical simulations of the electric field intensity enhancement inside a 110 nm ZMW respective to the homogeneous water reference, for (c) the 557 nm donor excitation, (d) the 635 nm acceptor excitation and (e) the 590 nm donor dipole radiation. The 3D shape of the ZMW is deduced from cross-cut SEM views of the fabricated sample. In the case of (c-d) the illumination is a 600 nm diameter spot incoming from the bottom of the image, while for (e) the position of the dipolar source is indicated by the cross and the contributions from two vertical and horizontal polarizations are averaged.

## Results and Discussion

The zero-mode waveguides are fabricated by focused ion beam (FIB) milling in opaque 100 nm thick aluminum films deposited on glass coverslips. Figure 1a shows typical SEM images of the fabricated ZMWs with diameters ranging from 85 to 170 nm. Ensuring the best optical performance for the ZMWs requires specific attention during the metal coating process, as aluminum is highly sensitive to the residual trace amount of oxygen found in the evaporation chamber.[64,65] Depending on the deposition parameters, the amount of oxide found within the bulk of the aluminum layer can dramatically change, affecting the dielectric permittivity and the plasmonic losses.[64,66] We have observed similar trends, and in order to reach the highest enhancement factors, we found that the aluminum deposition parameters critically need to reach the lowest chamber pressure (< $10^{-6}$ mbar) together with fast deposition rates (> 10 nm/s). FIB milling then accurately controls the final geometry of the ZMW allowing to optimize both the diameter and the 50 nm undercut in the glass below the aperture.[67–70]

The sketch of our experiment is depicted on Fig. 1b. The ZMWs are covered with the buffer solution containing the FRET constructs, leaving single molecules to freely diffuse at a concentration low enough to clearly isolate single FRET pairs inside the attoliter ZMW volume. Importantly to avoid unwanted adsorption of the molecules on the metal and glass surfaces, the zero-mode waveguides are passivated with a silane-modified polyethylene glycol (PEG-Silane, Fig. 1b).[71]

A critical issue at low FRET efficiencies is to discriminate between the molecular constructs featuring both a donor and an acceptor from those incompletely labelled lacking one fluorophore (or whose fluorophore is in a long-lived dark state). This issue is best dealt with using two alternating laser excitations to excite sequentially the donor and the acceptor dyes and ensure that for each detected donor there is a fluorescent acceptor on the molecular construct.[72,73] Here, we illuminate the ZMWs with pulsed interleaved excitation (PIE-FRET) featuring two alternating 557 and 635 nm picosecond laser pulse trains, each pulse being separated from the previous one by 12.5 ns.[73] Together with time-tagged time-resolved (TTTR) fluorescence detection, this PIE-FRET approach crucially selects only the FRET detection events and avoid the issues related to incomplete labelling. Otherwise in the FRET histograms, it would be very difficult in our case to separate the contribution from the FRET sample from one stemming from samples labelled only with the donor dye.

A major interest of aluminum as compared to gold is that it maintains good optical properties over the full visible spectrum,[65] and is thus well suited for multicolor laser excitation. Numerical simulations of the excitation profiles inside the ZMW (Fig. 1c,d) indicate that similar detection volumes and enhancement factors can be reached for both 557 and 635 nm laser wavelengths. This is an important feature for PIE-FRET experiments where the two beams have to be spatially overlapped. In the case of the donor dipole emission leading to FRET (Fig. 1e), the spatial profile for the enhancement of the FRET rate has a more complex shape, which depends on the donor position and orientation inside the ZMW (Supporting Information Fig. S1). While this complexity calls for experimental investigations, two global trends can still be extracted from the simulation in Fig. 1e. First, the FRET rate constants can be significantly enhanced inside the ZMW by a few fold as compared to the homogeneous environment reference. Second, the FRET rate enhancement increases for larger distances from the donor source, indicating that FRET efficiencies can be more improved at larger acceptor separations from the donor. This is exactly the configuration looked after to improve the FRET detection range well beyond the classical Förster radius. While this trend was previously inferred for plasmonic nanostructures,[39,46] and recently confirmed in the microwave regime,[24] its quantitative influence on extending the FRET range was not established so far.

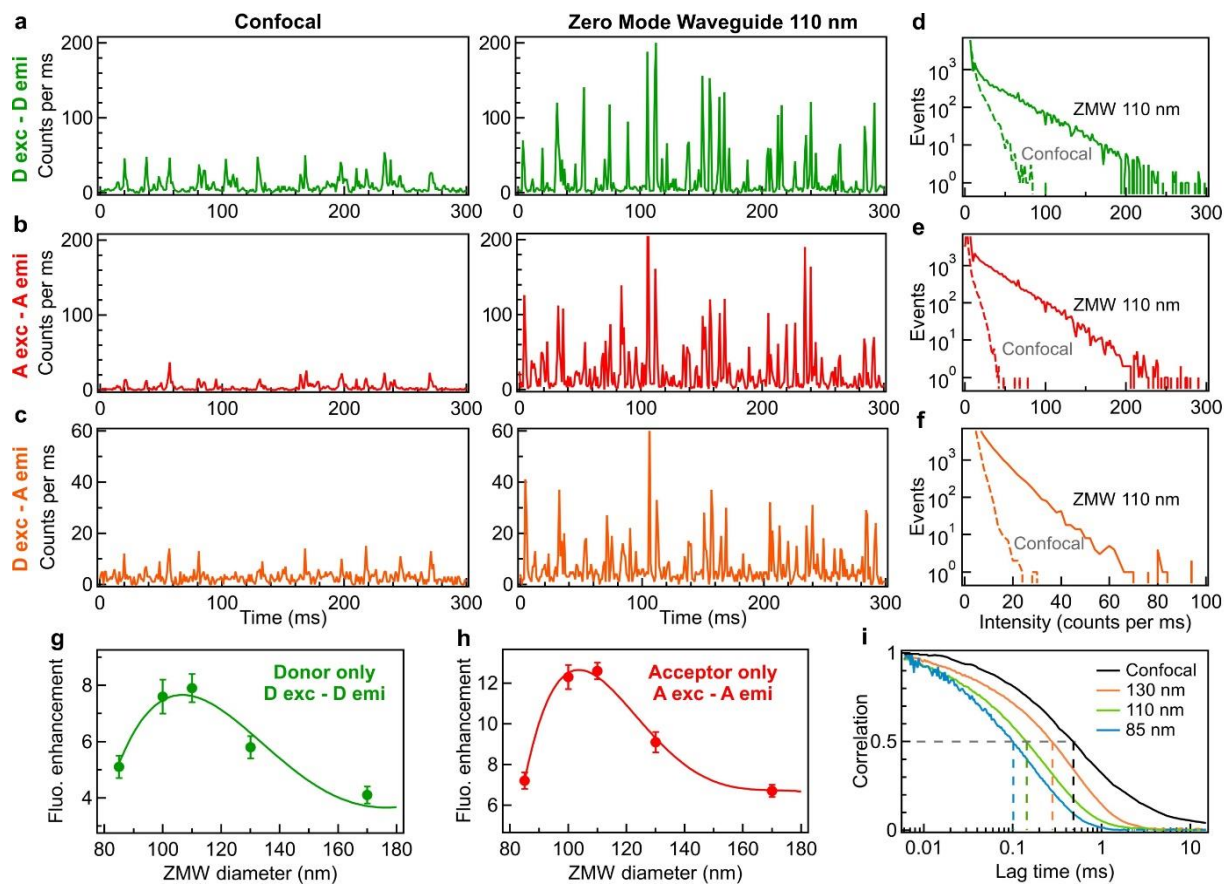

**Figure 2.** Enhanced fluorescence detection of single FRET pairs in zero-mode waveguides. (a-c) Fluorescence time traces recorded for the confocal reference and a 110 nm diameter ZMW in the cases of (a) donor emission after donor excitation at 557 nm, (b) acceptor emission after acceptor excitation at 635 nm, (c) acceptor emission after donor excitation at 557 nm (FRET case). The sample consists of single Atto550-Atto647N FRET pairs covalently linked to dsDNA with 13.6 nm (40 base pairs) separation between the dyes. The binning time is 1 ms. (d-f) Photon count histograms deduced from the full 60 s fluorescence traces in (a-c). (g-h) Fluorescence brightness enhancement measured by FCS as a function of the ZMW diameter for (g) the donor emission excited at 557 nm and (h) the acceptor emission excited at 635 nm. (i) Normalized FCS correlation functions for different ZMW diameters and for the diffraction-limited confocal reference, a clear reduction of the correlation function widths at half-maximum (dashed lines) confirms shorter diffusion times and smaller detection volumes in the ZMWs.

We now turn to the experimental characterization of PIE-FRET inside zero-mode waveguides. The FRET sample consists of one Atto550 molecule as donor (D) and one Atto647N as acceptor (A), with both dyes covalently attached to double stranded DNA molecules. The separation between the fluorophores is set by the DNA design to 40 base pairs, which corresponds to a donor-acceptor separation of about 13.6 nm, more than two times the 6.26 nm Förster radius for this FRET pair.[4] Figure 2a-c shows typical fluorescence time traces recorded with the confocal geometry and with a 110 nm diameter ZMW. Individual fluorescence bursts stemming from single molecules are clearly resolved when the molecules diffuse across the detection volume, with a strikingly higher brightness in the ZMW case as compared to the confocal case. This is especially important for detecting FRET counts (Fig. 2c): for 13.6 nm D-A separation, the FRET signal in the confocal case is of a few counts per millisecond, which is very close to the noise threshold. The situation is clearly different inside the ZMW where intensities higher than 30 counts/ms are readily detected. Additional time traces for different ZMW diameters are shown in the Supporting Information Fig. S2. We stress that all the data reported here are spatially averaged over the attoliter ZMW volume. The electric field distribution inside this volume is the main parameter determining the fluorescence enhancement, and the contribution from the edge plasmon modes located at the aperture rim remain marginal.[74,75]

The photon count histograms (Fig. 2d-f) from the full 60 s fluorescence traces confirm the brighter single-molecule detection events in the ZMW case. To accurately measure the fluorescence enhancement factors for both donor and acceptor dyes taken individually, we repeat this experiment for molecular samples labelled with either only the donor or only the acceptor (this calibration is also needed to quantify clearly the FRET efficiency, see Methods section for details). The time traces are analyzed by fluorescence correlation spectroscopy (FCS) to measure the average fluorescence brightness per molecule and compute the fluorescence enhancement factor in the ZMW.[74] A description of the FCS fitting procedure and non-normalized FCS data are presented in section S3 and Fig. S3 of the Supporting Information. Figure 2g,h summarizes our results on the fluorescence enhancement. A clear optimum diameter is found around 110 nm, providing enhancement factors of 8× for the donor and 12× for the acceptor. These values are the brightest reported so far with circular aluminum ZMWs for high quantum yield dyes such as Atto550 and Atto647N,[40,75,76] and could only be achieved after careful optimization of the aluminum deposition and FIB milling procedures. Additionally, the FCS analysis also shows that the DNA molecules are freely diffusing across the ZMW volume and that our measurements are not influenced by unspecific adhesion to the metal or glass surfaces (Fig. 2i).

We next follow the protocol established in Ref [4] to accurately quantify the FRET efficiency $E_{FRET}$ from the fluorescence bursts. First, PIE-FRET selection is applied to select only the events featuring both a donor and an acceptor. In brief, a first threshold is set on the sum of the detected photons in the donor and acceptor channels to separate the single molecule florescence bursts from the background noise. A second threshold is then applied to select the events where a signal is recorded in the acceptor channel upon red laser excitation, confirming the presence of the active acceptor dye (see Methods section for details).[63,73] Then the detection count rates are corrected to account for several additional effects, including (i) donor emission leakage (crosstalk) into the acceptor channel, (ii) direct excitation of the acceptor by the 557 nm laser, and (iii) differences in the quantum yield and detection efficiencies between the donor and acceptor emissions (see details in the Methods section). The presence of the ZMW modifies these corrections parameters as compared to the confocal case, therefore we have carefully characterized and calibrated each correction parameter for each specific ZMW diameter. We have also checked that the experimental uncertainties in the determination of these correction parameters do not influence our main scientific conclusions, which remain valid even if the different parameters used to compute the FRET histograms are varied (see complete details in the Supporting Information Fig. S4-S10).

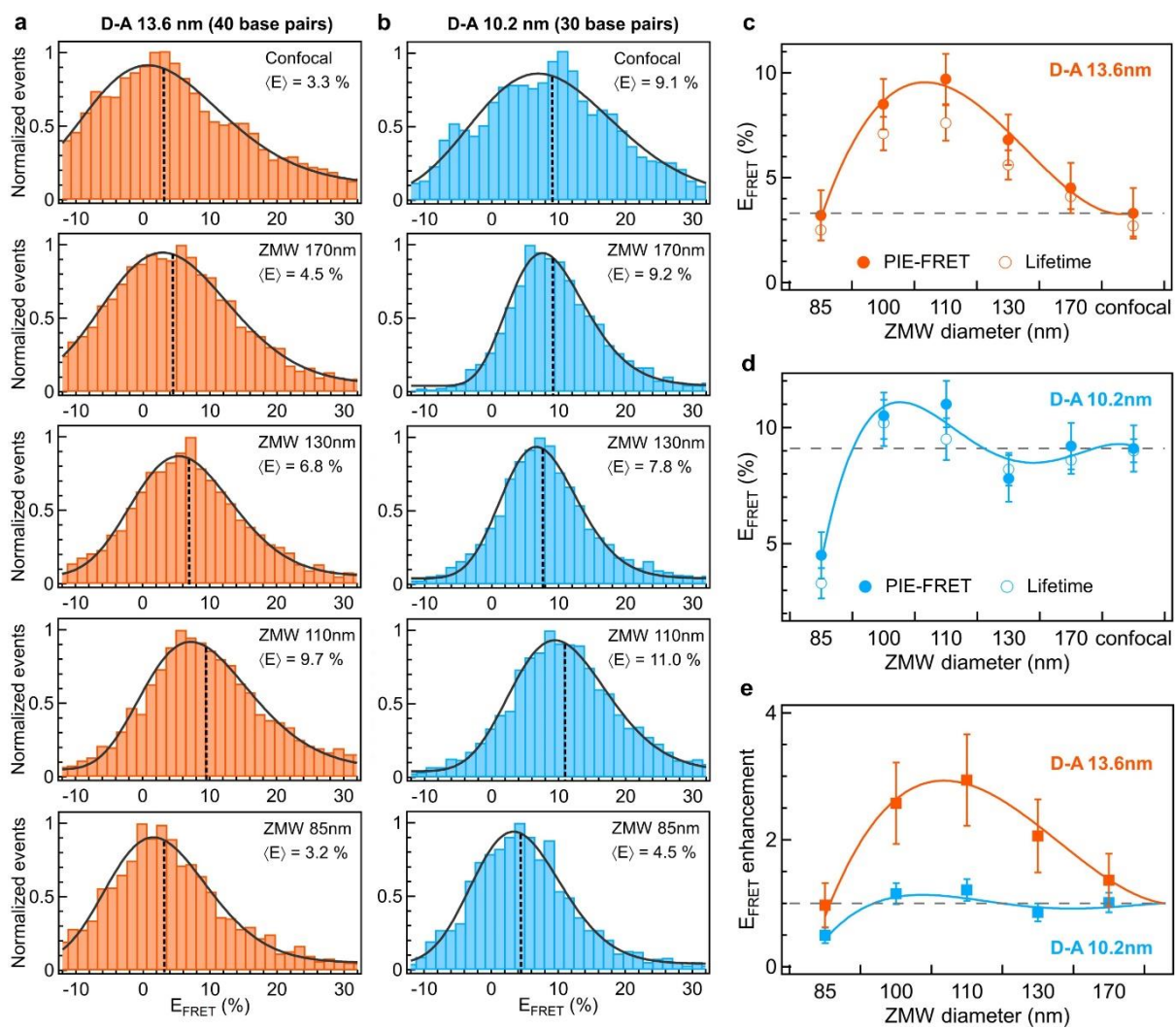

**Figure 3.** FRET efficiency enhancement with zero-mode waveguides. (a,b) Single molecule FRET histograms measured in ZMWs of different diameters and in the confocal reference for donor-acceptor separations of 13.6 nm (a) and 10.2 nm (b) respectively. Black lines are fits following a gamma distribution used to determine the average FRET efficiency indicated on each graph (see Methods section for details). (c) Average FRET efficiency as a function of the ZMW diameter for 13.6 nm donor-acceptor separation. Filled markers are deduced from the FRET histograms in (a) while empty markers are deduced from the donor fluorescence lifetime analysis (see Supporting Information section S10). The horizontal dashed line indicates the confocal reference and the solid line is a guide to the eyes. (d) Same as (c) for 10.2 nm donor-acceptor distance. (e) FRET efficiency enhancement deduced from (c,d) as compared to the confocal reference.

Figure 3a shows the single molecule FRET histograms obtained for the 13.6 nm D-A separation with different ZMWs. As the ZMW diameter is reduced from 170 to 110 nm, the average FRET efficiency shows a pronounced shift towards higher values, indicating energy transfer enhancement. Importantly, while for this large D-A distance the average FRET efficiency in the confocal case is only 3.3 %, it can be improved up to 9.7 % inside a 110 nm ZMW (Fig. 3c,e). Moreover, thanks to the brighter single molecule fluorescence counts (Fig. 2g,h), the statistical accuracy is also improved and the FRET histogram width is reduced in the ZMW. As pointed out by Deniz and coworkers,[77,78] a major contribution to the width of the smFRET histograms comes from the shot noise. Increasing the fluorescence brightness per pulse thus reduces the shot noise contribution by a factor corresponding to the square root of the fluorescence enhancement, leading to a pronounced reduction of the histogram widths.

If we keep on decreasing the ZMW size, the FRET efficiency enhancement is lost for the smallest 85 nm diameter ZMW. This observation is consistent with the reduction of the fluorescence brightness seen for the 85 nm ZMW (Fig. 2g,h), and is also a clear indication for the competition between the FRET process and the losses into the metal. As the ZMW diameter is reduced, the dyes come in closer proximity with the metal layer, leading to a larger contribution of the Ohmic losses into the metal which tend to quench the fluorescence and the FRET enhancement. As supplementary validity proof of our results, let us stress that the measurements of the average FRET efficiency based on the fluorescence lifetime confirm all the results obtained from the fluorescence burst analysis (empty markers in Fig. 3c,d, see Supporting Information section S10 for details). Our fluorescence lifetime measurements also show that the donor lifetime in the ZMW is clearly reduced by the presence of the acceptor at 13.6 nm distance (Supporting Information Fig. S12). This is a pure signature of FRET, highlighting the fundamental difference with far-field radiative energy transfer.[53–57]

In a second set of experiments, we investigate dsDNA constructs with 10.2 nm D-A separation (Fig. 3b,d). Although the photonic environment and the fluorescence enhancement are the same as before and we use exactly the same analysis procedure, here the average FRET efficiency is only moderately enhanced from 9.1 % in the confocal configuration to 11 % in the optimum 110 nm ZMW. Again, we find that the 85 nm ZMW leads to a significant quenching of the FRET efficiency. For D-A separations shorter that 10 nm, the ZMW does not bring a sufficient improvement on the FRET rate to overcome the other radiative and non-radiative decay pathways which are also affected by the ZMW. This goes in agreement with the numerical simulations (Fig. 1e) and with earlier works on FRET in ZMWs at short D-A separations.[40,41,43] The main result here is that for distances greater than 10 nm, the FRET process can be significantly enhanced inside an optimized ZMW, leading to a clear improvement of the FRET detection range.

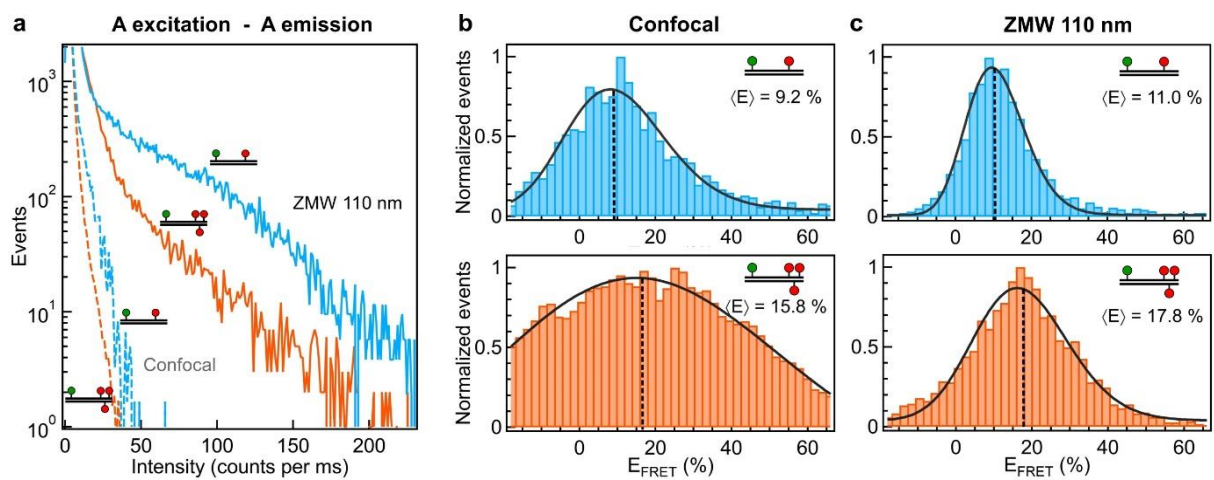

**Figure 4.** Multi-acceptor approach to further enhance the FRET efficiency. (a) Comparison of the photon count histograms for the acceptor emission after direct excitation at 635 nm for a 110 nm diameter ZMW (solid lines) and the confocal reference (dashed lines), in the cases of single acceptor (blue traces) and multiple acceptors (orange traces). (b,c) FRET histograms measured in the confocal reference (b) and in a 110 nm ZMW (c) for single and multi-acceptors. Black lines are fits following a gamma distribution used to determine the average FRET efficiency.

An alternative approach to extend the FRET detection range uses multiple acceptors to maximize the probability of donor energy transfer. [13–15] Interestingly, this approach can be combined with the ZMWs to maximize the gains and benefit from high photon count rates (Fig. 4). To test this combination, we use double stranded DNA constructs featuring a single Atto550 donor and three Atto647N acceptors located respectively at 31, 34 and 37 base pairs separation from the donor (corresponding to a minimal donor-acceptor separation of 10.6 nm). As the acceptors are separated by only 3 base pairs between them, self-quenching between the acceptors tends to reduce the acceptor fluorescence brightness (Fig. 4a and Supporting Information Fig. S17-S19). A recent systematic study using DNA origami to position several Atto 647N dyes independently confirms this observation.[79] In the case of ZMWs, the negative effect of self-quenching can be compensated by the fluorescence enhancement to provide single molecule detection events exceeding 50 photons per ms, well above the experimental detection limit (Fig. 4a). Again, we carefully calibrate the experimental parameters to compute the PIE-FRET histograms for both the confocal and the ZMW case. Figure 4b,c summarizes our results, and allows to directly compare the influence of using multiple acceptors, the ZMW, or both techniques. In the confocal geometry, the use of multiple acceptors improves the average FRET efficiency from 9.2 to 15.8%. This value is in excellent agreement with the independent calibration of multiple acceptors FRET in Ref. [13], which concluded to an enhancement of the apparent FRET transfer rate of $3^{0.58} \approx 1.9$, leading to a predicted FRET efficiency of 15.8%, which corresponds nicely to our experimental value. However in the confocal configuration, the self-quenching between neighboring acceptors leads to low photon count rates, which translates into broad PIE-FRET histograms. Fortunately, this situation can be further improved with the ZMW. Thanks to the high photon count rates in the ZMW, the FRET signal is more easily detectable well above the background noise and the FRET histogram becomes significantly narrower (Fig. 4c). Moreover, thanks to the optical confinement inside the ZMW and the FRET enhancement at long distances, we even monitor an additional increase of the average FRET efficiency up to 17.8 % in the 110 nm ZMW. This gain may seem moderate, but observing an improvement of the FRET efficiency is already an important result in the context of FRET with plasmonics, where most often a significant loss of the FRET efficiency is monitored.[23,25,31,36,37,45–47] Using a combination of multiple acceptors with optimized ZMWs takes maximum advantage of the enhancement of the FRET efficiency and the fluorescence signal to further ease the smFRET detection at long distances.

To discuss the physics behind our observations, we need to introduce the different donor decay rate constants as illustrated in Fig. 5a. From its excited state, the donor molecule can decay to the ground state *via* different radiative or nonradiative pathways. In a homogeneous medium, the donor decays radiatively with a rate $\Gamma_{rad}^0$, the rate for nonradiative internal conversion is $\Gamma_{nrad}^0$ and the FRET rate is $\Gamma_{FRET}^0$. This FRET rate is described by the classical Förster's formalism as $\Gamma_{FRET}^0 = (\Gamma_{rad}^0 + \Gamma_{nrad}^0)(R_0/R)^6$, where $R_0$ is the Förster radius and $R$ is the donor-acceptor distance. With these definitions the FRET efficiency in a homogeneous medium is defined as

$$E_{FRET}^0 = \frac{\Gamma_{FRET}^0}{\Gamma_{FRET}^0 + \Gamma_{rad}^0 + \Gamma_{nrad}^0} = \frac{1}{1 + (R/R_0)^6} \qquad (1)$$

In presence of the ZMW, the radiative decay rate is increased to $\Gamma_{rad}^{ZMW}$ as part of the radiated power is scattered by the ZMW back to the donor dipole (Purcell effect).[80] We assume that the nonradiative internal conversion rate $\Gamma_{nrad}^0$ is unchanged and we introduce a supplementary decay rate $\Gamma_{loss}^{ZMW}$ to account for the energy nonradiatively lost into the free electron cloud present in the metal. Regarding the energy transfer to the acceptor dipole, we can always write that the total energy transfer rate $\Gamma_{FRET}^{tot}$ is the sum of the FRET rate in a homogeneous medium $\Gamma_{FRET}^0$ plus an additional term $\Gamma_{FRET}^{ZMW}$ to account for the dipole emission backscattered to the acceptor position by the ZMW:

$$\Gamma_{\text{FRET}}^{\text{tot}} = \Gamma_{\text{FRET}}^{0} + \Gamma_{\text{FRET}}^{\text{ZMW}} \tag{2}$$

The physical origin behind this equation and the expression of $\Gamma_{\text{FRET}}^{\text{ZMW}}$ are detailed in the Supporting Information Section S14. Equation 2 shows that in the ZMW, the evolution of the total FRET rate can deviate from the classical $1/R^6$ distance dependence, as the ZMW contribution $\Gamma_{\text{FRET}}^{\text{ZMW}}$ follows a non-trivial distance dependence with the donor-acceptor distance due to the complex spatial distribution of the electric field radiated by the donor inside the ZMW (Fig. 1e). Using these definitions, we can express the FRET efficiency inside the ZMW as

$$E_{\text{FRET}}^{\text{ZMW}} = \frac{\Gamma_{\text{FRET}}^{0} + \Gamma_{\text{FRET}}^{\text{ZMW}}}{\Gamma_{\text{FRET}}^{0} + \Gamma_{\text{FRET}}^{\text{ZMW}} + \Gamma_{\text{rad}}^{\text{ZMW}} + \Gamma_{\text{loss}}^{\text{ZMW}} + \Gamma_{\text{nrad}}^{0}} \tag{3}$$

Observing an enhancement of the FRET efficiency $E_{\text{FRET}}^{\text{ZMW}}$ inside the ZMW as compared to the confocal reference $E_{\text{FRET}}^{0}$ depends on a delicate balance between all the different decay rates.[47] Of primary importance is the fact that the FRET rate contribution mediated by the ZMW $\Gamma_{\text{FRET}}^{\text{ZMW}}$ is significant as compared to the direct FRET contribution in a homogeneous medium $\Gamma_{\text{FRET}}^{0}$. This can be achieved for large donor-acceptor separations beyond the Förster radius as $\Gamma_{\text{FRET}}^{0}$ vanishes, making the ZMW contribution $\Gamma_{\text{FRET}}^{\text{ZMW}}$ stand out more prominently. Qualitatively, this explains our observations of the FRET efficiency enhancement being more important for 13.6 nm D-A separations than for 10.2 nm (Fig. 3). Another important parameter is to ensure that the plasmonic loss rate $\Gamma_{\text{loss}}^{\text{ZMW}}$ remains moderate as compared to the FRET rates. If this is not the case (as for the 85 nm ZMW here or for aluminum nanoantennas[46]), the FRET efficiency will be quenched.

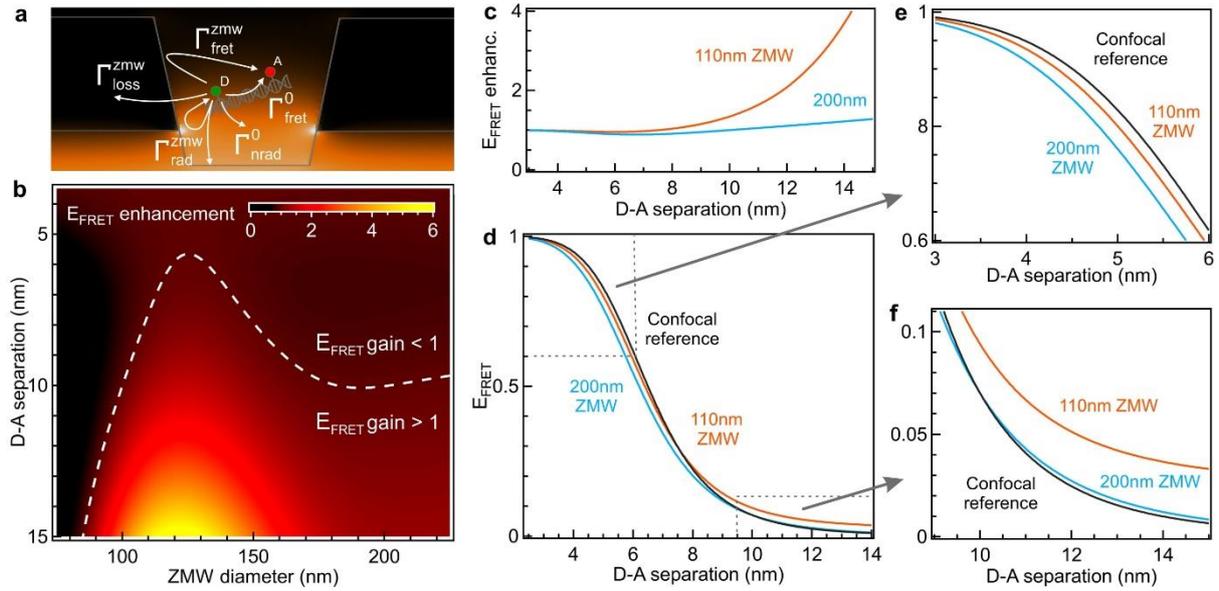

**Figure 5.** Guidelines for enhancing FRET with zero-mode waveguides. (a) Notations used to describe the different donor decay pathways. (b) FRET efficiency enhancement calculated as functions of the donor-acceptor separation and the ZMW diameter for the single Atto550 donor - single Atto647N acceptor construction. The dashed line indicates the contour where the FRET efficiency enhancement equals unity, separating the region where the FRET efficiency is enhanced (bottom) from the one where it is quenched (top). (c) FRET efficiency enhancement as a function of the donor-acceptor separation for two selected ZMW diameters. (d) Evolution of the average FRET efficiency as a function of the donor-acceptor separation assuming a 6.5 nm Förster radius in the confocal reference. Close-up views of the regions of short and large donor-acceptor separations are displayed in (e,f) respectively.

Building on the trends observed here and earlier studies at shorter D-A separations,[38–41] we derive a global map showing the FRET efficiency enhancement as functions of the donor-acceptor separation and the ZMW diameter (Fig. 5b). This map is intended to provide a global discussion of smFRET with ZMWs and further ease its future applications. Here the calculations assume the common case of a single Atto550-Atto647N donor-acceptor pair, and do not consider the more advanced constructions featuring multiple donors or acceptors. The computation of Fig. 5b is based on the evolution of the experimental donor fluorescence lifetime (Supporting Information Fig. S14). In the absence of the acceptor, it allows us to compute the total decay rate constant $\Gamma_{rad}^{ZMW} + \Gamma_{loss}^{ZMW} + \Gamma_{nrad}^{0}$ and interpolate its evolution with the ZMW diameter. The difference between the donor lifetime in presence of absence of the acceptor then allows us to compute the total FRET rate constant $\Gamma_{\text{FRET}}^{\text{tot}}$ and again interpolate its evolution with the ZMW diameter. We do this for several D-A distances shown in this work and also on previous studies on shorter separations,[38–41] and fit the evolution with the D-A distance. Then the calculation of the FRET efficiency follows Eq. 3. Lastly, the confocal reference FRET efficiency is calculated with Eq. 1 assuming a 6.5 nm Förster radius.

Different conclusions can be drawn from Fig. 5b. First, for D-A separations below 6-7 nm (typically the Förster radius in homogeneous space), the ZMW do not promote the FRET efficiency whatever the diameter is set to. For D-A distances below the Förster radius, the FRET rate $\Gamma_{\text{FRET}}^{0}$ dominates the other donor decay rates. In these cases, starting with a high FRET rate value, the ZMW does not improve sufficiently the FRET rate to yield a noticeable effect overcoming the other phenomena competing with FRET. As a result, the net FRET efficiency is reduced in the presence of the ZMW. However, the ZMW ability to improve the net fluorescence count rates (fluorescence enhancement) is preserved, so smFRET with high count rates can still be performed.

Second, for very small ZMW diameters below 90 nm, the quenching losses $\Gamma_{\text{loss}}^{\text{ZMW}}$ are dominating, and the FRET efficiency is reduced in the presence of the metal structure, independently of the D-A separation. Simultaneously, the ZMW ability to improve the detected fluorescence counts is lost. Therefore, ZMWs with diameters below 90 nm act very much as energy sinks, dissipating the donor's energy into heat. Apart for applications requiring extremely small zeptoliter volumes for single molecule detection, this zone should better be avoided for smFRET measurements.

Third, there exists a zone for D-A separations greater than 10 nm and ZMWs diameter between 100 and 150 nm where the FRET efficiency can be significantly enhanced. This constitutes the ideal zone to extend the smFRET measurement range with bright detection events well above the experimental noise. As the $E_{\text{FRET}}$ enhancement increases with the D-A separation (Fig. 5c), it seems appealing to work at very long distances exceeding 15 nm. However, for practical smFRET applications, the most important parameter is not the $E_{\text{FRET}}$ enhancement, but the net detected FRET efficiency.

In addition to discussing the $E_{\text{FRET}}$ enhancement, it is also important to display the evolution of the FRET efficiency as a function of the D-A separation (Fig. 5d-f). In the presence of the ZMW, this curve deviates from the $1/(1+(R/R_0)^6)$ formula derived with the classical Förster's theory in homogeneous space. For short D-A distances below 7 nm, we again find that the FRET efficiency in ZMWs is below the classical value in homogeneous space (black line in Fig. 5d), yet this decrease is of a few percent only, and can be partly compensated for the 110 nm optimum diameter as compared to the 200 nm diameter. For larger D-A distances above 9 nm, the 110 nm ZMM increases the detected FRET efficiency and makes smFRET more detectable above the experimental noise level. If we assume 2% to be the minimum detectable FRET efficiency, then the maximum FRET range using a 110 nm ZMW goes beyond 17 nm. It is also important to mention that for D-A distances above 13 nm, the FRET efficiency curve becomes relatively flat with the D-A separation (Fig. 5f). This means that while FRET is detectable at long distances, the accuracy of the distance measurements based on smFRET efficiency in ZMW is in the nanometer range for D-A distances above 13 nm. The price to pay for extending the smFRET range is a reduction in the sub-nanometer accuracy for determining distances at large D-A separations.

**Conclusions**

Because they offer high fluorescence signal-to-background ratio in attoliter volumes, zero-mode waveguides are appealing structures to detect single molecules at high micromolar concentrations with enhanced brightness.[60–62] Here we show that ZMWs can also enable the detection of single molecule FRET well beyond the classical 10 nm barrier with classical fluorescent dyes constructs. This significantly extends the distance range for smFRET measurements, enabling the possibility to monitor structures and conformational dynamics on bigger molecular constructs with enhanced brightness. While classical smFRET performs nicely below 10 nm and superresolution microscopy achieves spatial localization accuracy down to 20 nm, smFRET in ZMW bridges the gap between these two techniques, enabling distance measurements in the 10-20 nm range. The ZMW approach can also be combined with more advanced FRET constructs featuring several acceptors to further enhance the detected FRET efficiency. This constitutes a supplementary improvement to enable exploring what classical smFRET cannot see.

**Methods**

*Zero-mode waveguide fabrication*

Clean microscope glass coverslips are coated with a 100 nm-thick layer of aluminum deposited by electron-beam evaporation (Bühler Syrus Pro 710). To obtain the best optical performance for the aluminum layer, the chamber pressure during the deposition is set to levels below $10^{-6}$ mbar and the deposition rate is 10 nm/s. ZMWs are then milled into the aluminum layer using gallium-based focused ion beam (FEI dual beam DB235 Strata) with settings at 30 kV energy and 10 pA beam current.

*DNA samples*

A double stranded DNA oligonucleotide with 51 base pairs length was designed where the forward strand is labelled with Atto 550 (donor) and its complimentary reverse strand is labelled with Atto 647N (acceptor). Two DNA constructs were designed where donor (D) and acceptor (A) fluorophores are separated by 30 and 40 base pairs, providing D-A separations of 10.2 and 13.6 nm respectively. The sequence of the forward strand is 5'-CCT GAG CGT ACT GCA GGA TAG CCT ATC GCG TGT CAT ATG CTG T**T**C AGT GCG-3' where at position 44 the thymine (T) base is labelled by Atto 550 fluorophore. The complimentary reverse strand sequence is 5'-CGC ACT GAA CAG CAT ATG ACA CGC GAT AGG CTA TCC **T**GC AGT ACG C**T**C AGG-3'. For the DNA construct with 13.6 nm D-A separation, the T base of the reverse strand at position 47 is labelled by Atto 647N to obtain a 40 base pairs separation between donor and acceptor. For the sample with 10.2 nm separation, the T base at position 37 is labelled to have a 30 base pairs separation. The sequence of the multi acceptor reverse DNA strand is 5' CGC ACT GAA CAG CAT ATG ACA CGC GAT AGG CTA TCC TG**C** AG**T** AC**G** CTC AGG 3', where positions 39, 42 and 45 are labelled with Atto 647N fluorophore. As a result, in the multi acceptor double stranded DNA, the three A in the reverse strand are separated from the D in the forward strand by 31, 34 and 37 base pair respectively leading to a D-A separation distance of 10.5, 11.5 and 12.6 nm respectively.

All the HPLC purified DNA sequences were purchased from IBA life solution (Gottingen, Germany). The forward and reverse strands were annealed at 5 µM concentration in a buffer containing 5 mM Tris, 20 mM $MgCl_2$, 5 mM NaCl, pH 7.5 by first heating at 95°C for 5 minutes followed by a slow and stepwise cooling to room temperature. The reference D and A only double stranded DNA were prepared by annealing the D and A labelled oligonucleotide strand with its complimentary unlabeled DNA strand respectively. The double stranded DNA is diluted in 20 mM Hepes, 10 mM NaCl, 5% (v/v) tween 20 buffer to 100 pM and 100 nM for the smFRET measurements in confocal and in Al ZMW respectively. Hepes (≥ 99.5%, molecular biology grade), tris(hydroxymethyl)aminometane (Tris, ≥99.8%) NaCl, $MgCl_2$, tween 20 for the preparation of the experimental buffer were purchased from Sigma Aldrich and received without further purification.

*Surface passivation*

To avoid unwanted adsorption of the DNA FRET sample on the aluminum or glass surface, we passivate the ZMW surface with a silane-modified polyethylene glycol of molecular weight 1000 Da (PEG-silane 1000, Interchim). First, the ZMW sample is cleaned for 5 minutes by using an air plasma cleaner to remove any organic impurities. Immediately after plasma cleaning, the ZMW sample is covered with a solution of 1 mg/ml PEG1000-silane in absolute ethanol (≥ 99.7 %, Carlo Erba Reagent) with 1% acetic acid (AR grade, Sigma Aldrich) and left overnight at room temperature (20°C) under argon atmosphere to passivate the surface. Next the ZMWs are rinsed with ethanol to remove any excess unadsorbed PEG-silane and dried with a flow of synthetic air.

*Experimental setup*

All the smFRET measurements are performed in a home built confocal microscope set up with pulsed interleaved excitation.[73,81] The Atto 550 donor is excited at 557 nm by a iChrome-TVIS laser (Toptica GmbH, pulse duration ~ 3 ps). The Atto 647N acceptor is excited at 635 nm by using a LDH series laser diode (PicoQuant, pulse duration ~ 50 ps). The lasers are synchronized to operate at the same 40 MHz repetition rate, with a constant 12.5 ns delay between the green and red laser pulses. The set of alternating color laser pulses allows to temporally record the donor emission, the FRET signal and the acceptor emission. The two laser beams are spatially overlapped by using a dichroic mirror (ZT561RDC, Chroma), then they are reflected towards the microscope by a multiband dichroic mirror (ZT 405/488/561/640rpc, Chroma). The excitation power for both lasers is kept at 20 µW (measured at the microscope entrance port).

A Zeiss C-Apochromat 63x, 1.2 NA water immersion objective focuses the light on a single ZMW milled on an aluminum film. The donor and acceptor fluorescence are collected by the same objective in an epifluorescence configuration, and pass through the multiband dichroic (ZT 405/488/561/640rpc, Chroma) which separates the fluorescence from the laser backreflection. A supplementary emission filter (ZET405/488/565/640mv2, Chroma) is used to further suppress the laser back reflected light. The donor and acceptor fluorescence are spectrally separated into two detection channels with a dichroic mirror (ZT633RDC, Chroma). Each detection channel is equipped with a 50 µm pinhole and emission filters for spatial and spectral filtering of the fluorescence light. The donor channel is equipped with ET570LP and ET595/50m (Chroma) emission filters. The acceptor channel is equipped with ET655LP (Chroma) emission filter. Two single photon avalanche photodiodes (MPD-5CTC with < 50 ps timing jitter, Picoquant) are used to detect the donor and acceptor fluorescence. Each fluorescence photon is recorded with individual timing and channel information by a fast time correlated single photon counting module (HydraHarp400, PicoQuant) in a time tagged time resolved (TTTR) mode. Fluorescence lifetime measurements have a temporal resolution of 38 ps upon green excitation (557 nm) for and 110 ps upon red excitation (635 nm) defined as the full width half maximum of the instrument response function respectively.

*smFRET data analysis*

Low concentration of the DNA sample (100 pM for confocal and 100 nM for the nanohole) ensures that the fluorescence bursts stem from single molecules as the probability of having more than one molecule in the observation volume (femtoliter for confocal and attoliter for ZMW) is negligible. The fluorescence bursts are recorded in the donor and acceptor channels at a 1 ms binning time, which is close to the diffusion time of the DNA and is found to be optimal ensuring good signal to noise ratio and correct time resolution (Supporting Information Fig. S8 and S9). A first threshold criterion is applied on the sum of the detected photons in the donor and acceptor channels to select the single molecule florescence bursts for analysis and separate them from the background noise. In our case the threshold level is set at 25 counts per ms. A second threshold is then applied to select only the bursts indicating the presence of the acceptor dye upon red laser excitation. We use a threshold value of 12 photons per ms in the acceptor channel upon red excitation. For the confocal reference, we have to adapt the threshold values as the count rates are significantly lower. Hence we use 12 counts per ms for the threshold on the sum for burst detection and 3 counts per ms for the threshold on the acceptor channel upon red excitation. We carefully checked that the threshold levels used here do not influence the measured average FRET efficiencies.

The FRET efficiency $E_{FRET}$ is determined for each selected burst following the procedure commonly used in smFRET analysis.[4,81] Several phenomena must be taken into account in the $E_{FRET}$ calculation in order to get a reliable quantitative estimate. These phenomena include: (i) the leakage of the donor emission into the acceptor detection channel (crosstalk), (ii) the direct excitation of the acceptor emission by the green laser beam and (iii) the correction factor $\gamma$ to account for the differences in the fluorescence quantum yields ($\phi_D$, $\phi_A$) and detection sensitivities ($\kappa_D$, $\kappa_A$) of the donor and acceptor fluorophores. To determine the real number of photons due to FRET, the contributions due to the crosstalk ($\alpha$) and direct excitation ($\delta$) have to be subtracted from the detected fluorescence intensity in the acceptor channel. The corrected FRET efficiency is defined as

$$E_{FRET} = \frac{n_A^{green} - \alpha\, n_D^{green} - \delta\, n_A^{red}}{(n_A^{green} - \alpha\, n_D^{green} - \delta\, n_A^{red}) + \gamma\, n_D^{green}} \qquad (4)$$

where $n_D^{green}$ and $n_A^{green}$ are the numbers of photons per burst detected in the donor and acceptor channels following a green excitation pulse, and $n_A^{red}$ is the number of photons per burst detected in the acceptor channel following a red excitation pulse.

The crosstalk correction factor $\alpha$ is defined as the ratio of the donor emission intensity leaked into the acceptor channel as compared to that found in the donor channel following donor excitation: $\alpha = n_A^{green}/n_D^{green}\big|_{donor\ only}$. We calibrate $\alpha$ for each ZMW diameter and the confocal reference using the DNA sample labelled only with the donor dye. For the range of ZMW diameters probed here, we found that α is almost constant at 0.08, and moderately increased as compared to the α = 0.06 value found for the confocal reference. This is expected as the aluminum nanoapertures feature a broad spectral response covering the full donor emission spectra and thus do not significantly modify the donor emission spectrum.

The direct excitation is defined as the ratio of the acceptor emission intensity detected in the acceptor channel due to green excitation as compared to the acceptor emission intensity in the acceptor channel due to red excitation, when the sample is only labelled with the acceptor dye: $\delta = n_A^{green} / n_A^{red} \big|_{acceptor\ only}$. Here again, we carefully calibrate the δ correction factors for each ZMW diameter by performing experiments on DNA samples labelled only with the acceptor dye. For the confocal reference, δ amounts to 0.16. For 170 and 130 nm ZMWs, δ is measured to be 0.10, while for 110 and 100 nm diameters we find 0.08 and 0.075 respectively. For the smallest 85 nm size, δ increases again to 0.10. The evolution of δ scales as the ratio of green and red excitation intensities inside the ZMW, and varies slightly as the aperture diameter is changed, in agreement with the numerical simulations (Fig. 1c,d) and the experimental data (Fig. 2g,h).

The γ correction factor is defined as the ratio between the fluorescence quantum yields ($\phi_D$, $\phi_A$) and detection sensitivities ($\kappa_D$, $\kappa_A$) of the donor and acceptor fluorophores: $\gamma = \frac{\kappa_A \phi_A}{\kappa_D \phi_D}$. For the Atto 550 – Atto 647N FRET pair in the confocal configuration, we calculate $\gamma_{conf}$ = 0.80 ± 0.02 for our set of filters, in good agreement with the experimental determination using the stoichiometry S (see supporting information Table S6 and Fig. S16). In the presence of the ZMW, the γ parameter is modified due to the different enhancements of the quantum yields for the donor and acceptor. Its evolution can be quantified by the following equation:[39,46]

$$\gamma_{ZMW} = \gamma_{conf} \times \frac{EnhCRM_{AO}^{green}}{EnhCRM_{DO}^{green}} = \gamma_{conf} \times \frac{\delta_{ZMW}}{\delta_{conf}} \times \frac{EnhCRM_{AO}^{red}}{EnhCRM_{AO}^{green}} \qquad (5)$$

where $EnhCRM_{AO}^{green}$, $EnhCRM_{DO}^{green}$ are the fluorescence enhancement factors for count rate per molecule (CRM, or fluorescence brightness per molecule) of the acceptor-only and donor-only molecules following green excitation, and $EnhCRM_{AO}^{red}$ is the fluorescence enhancement factor for the acceptor-only molecules with red laser excitation. Our calibrations performed for each ZMW diameter quantify the $EnhCRM_{DO}^{green}$ and $EnhCRM_{AO}^{red}$ enhancement factors (displayed in Fig. 2g,h respectively) and also the evolution of the δ parameters, so the $\gamma_{ZMW}$ corrections factors can be computed for each ZMW. The $\gamma_{ZMW}$ determined for 170, 130, 110, 100 and 85 nm ZMWs are 0.71, 0.80, 0.70, 0.70 and 0.80 respectively, and are only slightly modified by the ZMW as compared to the confocal reference. We have also experimentally determined the $\gamma_{ZMW}$ correction factor using the measured photon stoichiometry S (Supporting Information Fig. S15), both estimates stand in excellent agreement for all the different ZMW diameters (Supporting Information Fig. S16), which further confirms the validity of our approach.

For the multi acceptor DNA FRET construct, $\alpha$ and $\delta$ are determined as 0.07 and 0.16 respectively for the confocal reference, and 0.077 and 0.10 in the case of the 110 nm ZMW. For the multi acceptor sample, the $\gamma$ value will be further modified from the single acceptor sample due to their different fluorescence brightness. For confocal reference, $\gamma_{multi}$ is determined as $\gamma_{multi} = \gamma_{single} \times CRM_{multi\,A}/CRM_{single\,A} = 0.26$. For the ZMW, the $\gamma_{multi}$ will be similarly modified due to different fluorescence enhancement for donor and multi acceptor sample as we observed in the case of the single acceptor constructs. Therefore, $\gamma_{multi}$ is determined from Eq. 2 to be 0.34.

*Fitting the FRET histograms*

The smFRET histograms are fitted here with a gamma distribution in order to better account for the asymmetric shape of the distribution between values higher than the median as compared to values lower than the median. We use the following model to fit the normalized number of events:

$$N_{events}(E) = \frac{A}{\Gamma(k)\theta^k} \, (E + E_0)^{k-1} \, e^{-(E+E_0)/\theta} + N_0 \qquad (6)$$

where E is the variable, A a scaling factor, k the shape parameter, $\theta$ the scale parameter, and $\Gamma(\ )$ is the mathematical gamma function. $N_0$ is a vertical offset to take into account some residual baseline noise, we find that $N_0$ never exceeds a few percents and is quite negligible. $E_0$ is an offset parameter needed to take into account the values where E is negative due to the noise distribution. As the gamma distribution is only defined for positive variables, we have to use this offset so that the fitting converges. The contribution of $E_0$ is of course subtracted to estimate the average FRET efficiency, which is given by:

$$\langle E_{FRET} \rangle = k \, \theta - E_0 \qquad (7)$$

A comparison with a Gaussian fit is provided in the Supporting Information Fig. S10. While the gamma distribution provides a nicer interpolation to the data, the Gaussian approach remains largely valid, and does not modify any of our conclusions.

*Numerical simulations*

Computations for the electric field distributions inside the ZMW are performed with finite-difference time-domain (FDTD) method using RSoft Fullwave software. In all the simulations, the size and shape of the ZMW are set to reproduce our actual experimental samples based on the SEM imaging (FEI dual beam DB235 Strata). The complex permittivity for aluminum is taken from the optimized experimental values recorded in ref [66]. The substrate refractive index is set to 1.52 and corresponds to borosilicate glass coverslips. The ZMW inner volume and top space are filled with water (refractive index 1.33). Each simulation is run with 1 nm mesh size and is checked for convergence after several optical periods. The spatial maps presenting the enhancement of the energy transfer rate (Fig. 1e) are calculated as the ratio of the field intensity distribution $|E_D(r_A)|^2$ generated by the donor in presence of the ZMW or in homogeneous space of refractive index 1.33. The ratio $|E_D(r_A)|^2_{ZMW}$ / $|E_D(r_A)|^2_{free\ space}$ directly corresponds to the increase in the energy transfer rate from the donor to the acceptor, as demonstrated in ref [80].

**Supporting Information**

Orientation-dependent numerical simulations of the donor intensity inside the ZMW, Single molecule fluorescence time traces with ZMWs of different diameters, FCS analysis with ZMWs, FRET histograms recorded for the sample labelled only with the donor, Influence of the crosstalk parameter $\alpha$ on the measured FRET efficiency, Influence of the direct excitation parameter $\delta$ on the measured FRET efficiency, Influence of the $\gamma$ parameter on the measured FRET efficiency, Influence of the binning time on the measured FRET efficiency, Comparison of models to fit the smFRET histograms, Fluorescence lifetime analysis, S-E plot diagrams, Experimental determination of the $\gamma$ correction factor, Fluorescence spectra of the multi-acceptor DNA sample, Expression of the total energy transfer rate constant inside the ZMW.

**Acknowledgments**

The authors thank Emmanuel Margeat for helpful discussions and Satish Moparthi for help with the DNA sample design and preparation. This project has received funding from the European Research Council (ERC) under the European Union's Horizon 2020 research and innovation programme (grant agreement No 723241) and from the Agence Nationale de la Recherche (ANR) under grant agreement ANR-17-CE09-0026-01.

**Supporting Information for**

**Extending Single Molecule Förster Resonance Energy Transfer (FRET) Range**

**Beyond 10 Nanometers in Zero-Mode Waveguides**

Mikhail Baibakov,[*] Satyajit Patra,[*] Jean-Benoît Claude, Antonin Moreau, Julien Lumeau, Jérôme Wenger[#]

*Aix Marseille Univ, CNRS, Centrale Marseille, Institut Fresnel, 13013 Marseille, France*

[*] *These authors contributed equally to this work*

[#] *Corresponding author: jerome.wenger@fresnel.fr*

**Contents:**

**S1. Orientation-dependent numerical simulations of the donor intensity inside the ZMW**

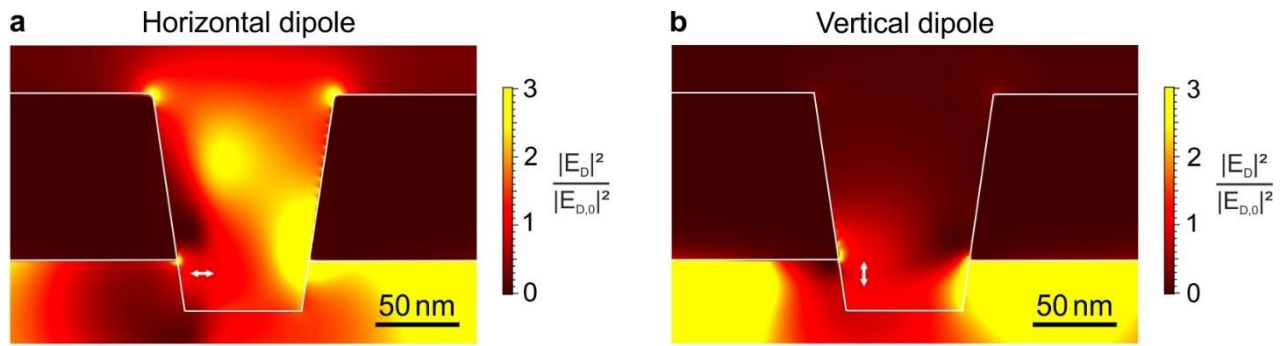

**Figure S1**: Numerical simulations of the electric field intensity enhancement $|E_D(r_A)|^2{}_{ZMW}$ / $|E_D(r_A)|^2{}_{free\ space}$ inside a 110 nm ZMW respective to the homogeneous water reference for the 590 nm donor dipole radiation. The source dipole is located at the position of the arrow with its orientation indicated on the graph.

## S2. Single molecule fluorescence time traces with ZMWs of different diameters

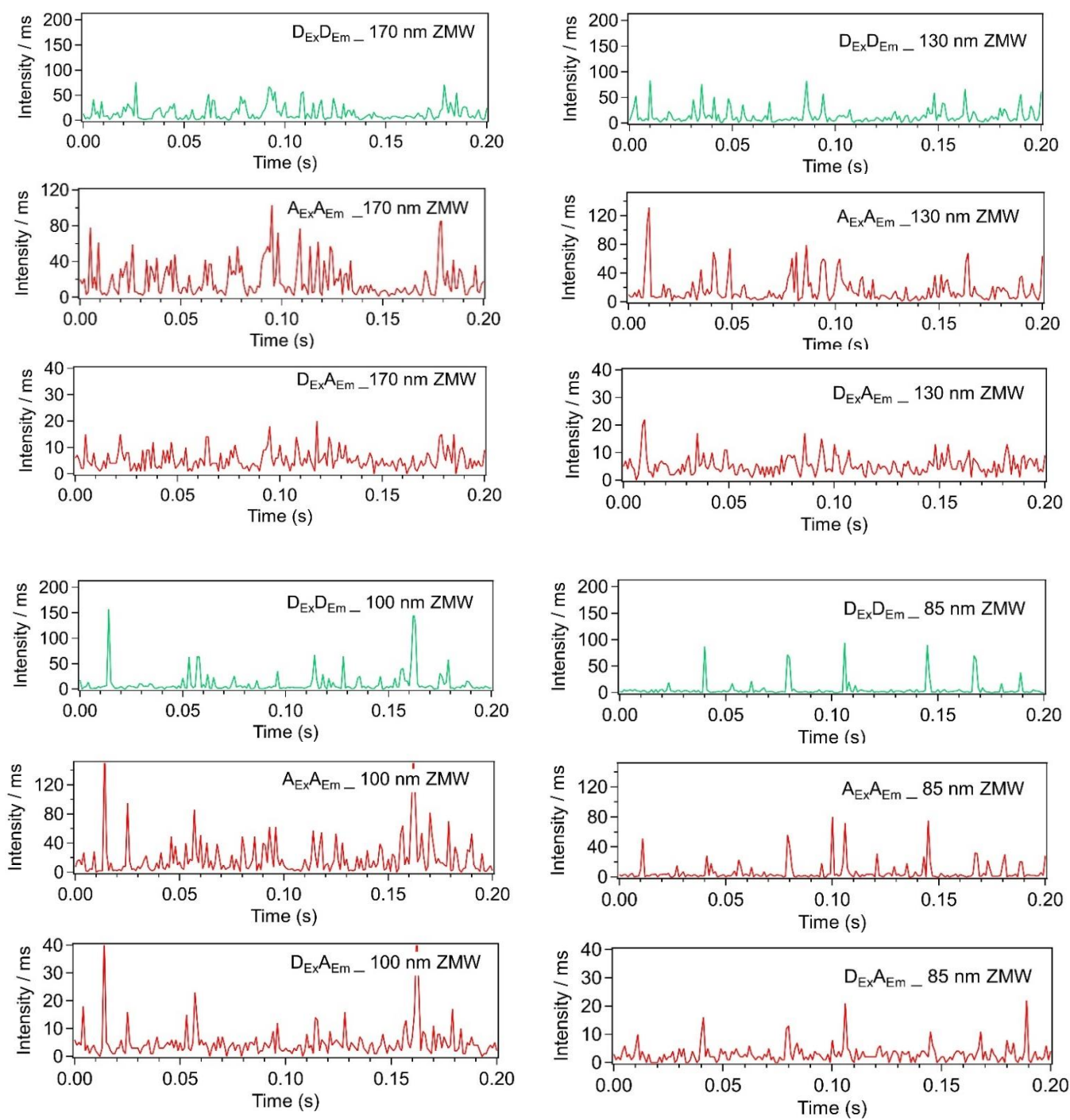

**Figure S2**: Typical PIE-FRET single molecule fluorescence intensity time traces obtained using ZMWs of diameters 170, 130, 100 and 85 nm respectively. $D_{Ex}D_{Em}$ stands for green (donor) excitation, donor emission, $A_{Ex}A_{Em}$ stands for red (acceptor) excitation, red emission, and $D_{Ex}A_{Em}$ stands for green (donor) excitation, acceptor emission (FRET signal).

## S3. FCS analysis in ZMWs

The FCS correlation data is fitted using a three dimensional Brownian diffusion model with an additional blinking term: [S1]

$$G(\tau) = \frac{1}{N}\left[1 + \frac{T_{ds}}{1-T_{ds}}\exp\left(-\frac{\tau}{\tau_{ds}}\right)\right]\left(1+\frac{\tau}{\tau_d}\right)^{-1}\left(1+\frac{1}{\kappa^2}\frac{\tau}{\tau_d}\right)^{-0.5} \tag{S1}$$

where N is the total number of molecules, $T_{ds}$ the fraction of dyes in the dark state, $\tau_{ds}$ the dark state blinking time, $\tau_d$ the mean diffusion time and $\kappa$ the aspect ratio of the axial to transversal dimensions of the nanohole volume. Figure S3 shows representative FCS correlation functions and their numerical fits for different ZMW diameters. While the ZMW geometry obviously does not fulfill the assumption of free 3D diffusion, the above model equation was found to empirically describe well the FCS data inside ZMWs, provided that the aspect ratio constant is set to $\kappa = 1$ as found previously. [S2, S3] The only residual difference is the long tail of the FCS function two times longer than the diffusion time, which is not so well accounted for by this model. Nevertheless, estimating relative ratios between the number of molecules (given by the FCS amplitude) and diffusion time (given by the FCS width) remains always possible.

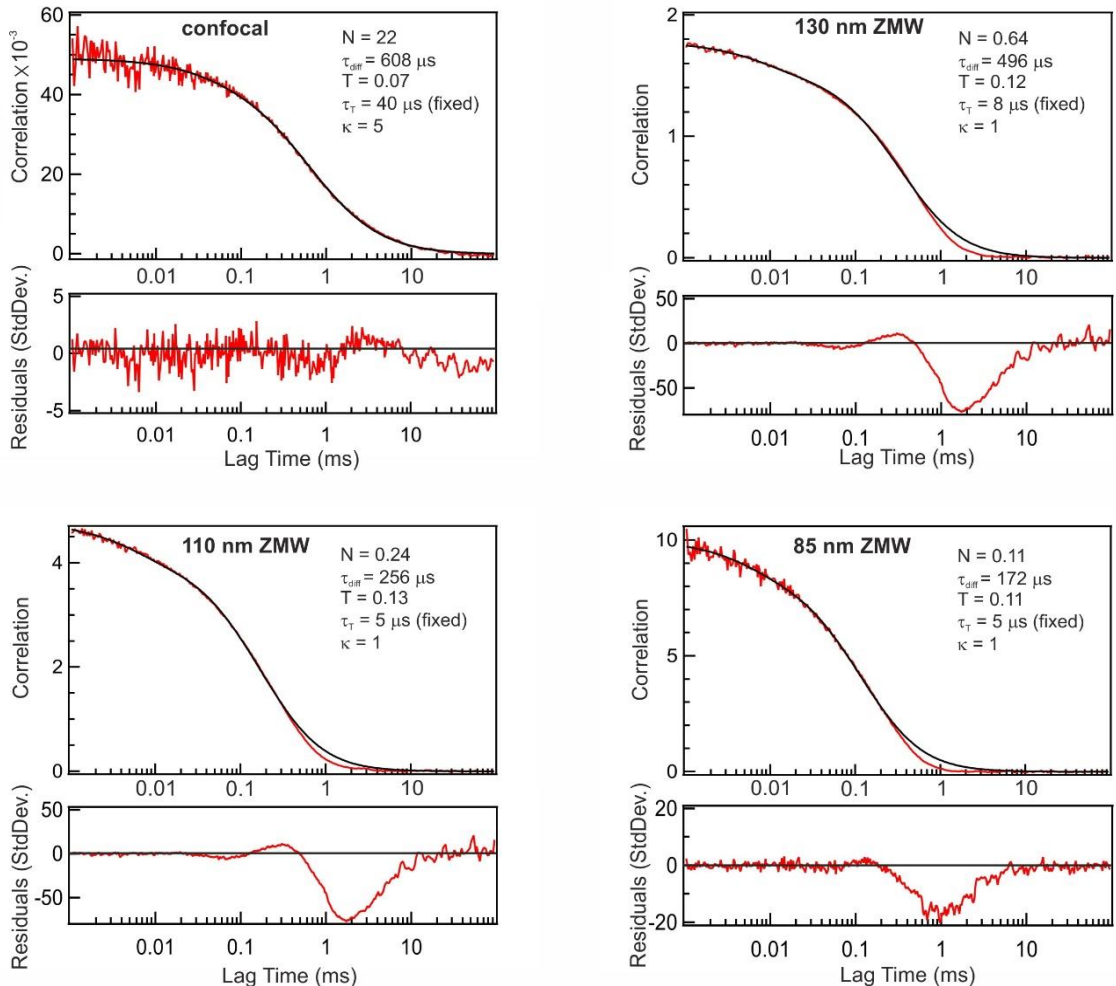

**Figure S3**: FCS correlation (red curves) and numerical fits (black) following the model in Eq. (S1) for the confocal reference and ZMWs of three different diameters. The lower trace shows the residuals from the fit. The parameters deduced from the fitting are indicated each panel. The DNA concentration used for the ZMWs is 100 nM, and 20 nM for the confocal experiment shown here.

## S4. FRET histograms recorded for the sample labelled only with the donor (zero FRET reference)

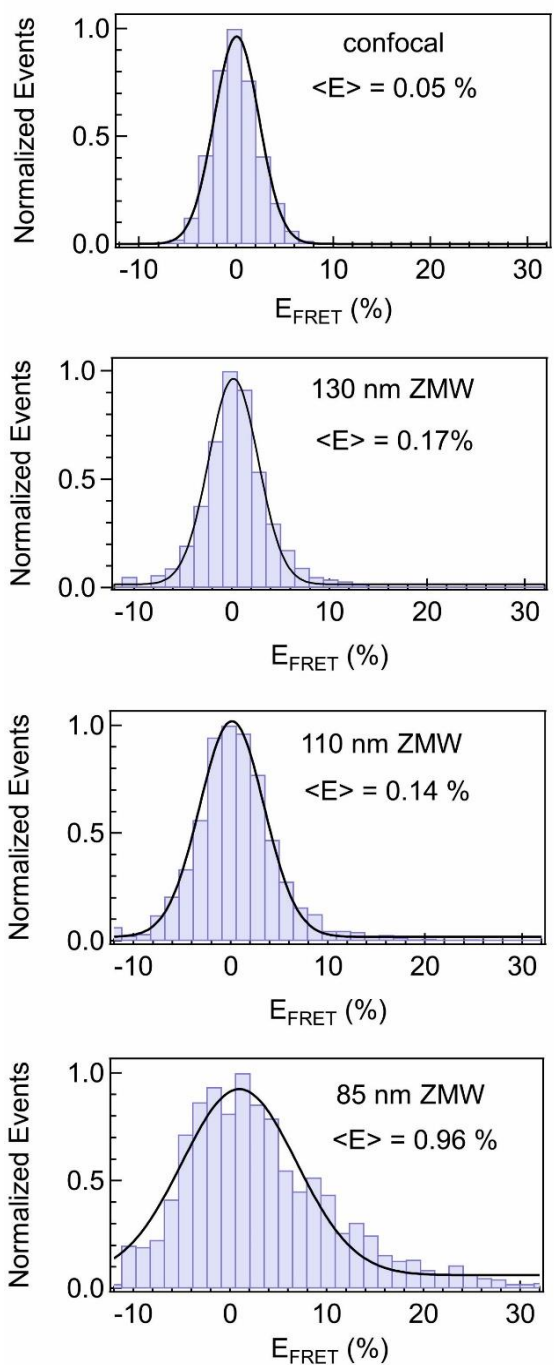

**Figure S4**: FRET efficiency histograms recorded for the dsDNA labelled only with the donor dye, in the confocal setup and in presence of ZMWs of different diameters. We find that for all the cases, the donor-only histograms feature a Gaussian distribution centered around 0%, which clearly differ from the histograms recorded for the FRET samples (Fig. 3).

## S5. Influence of the crosstalk parameter α on the measured FRET efficiency

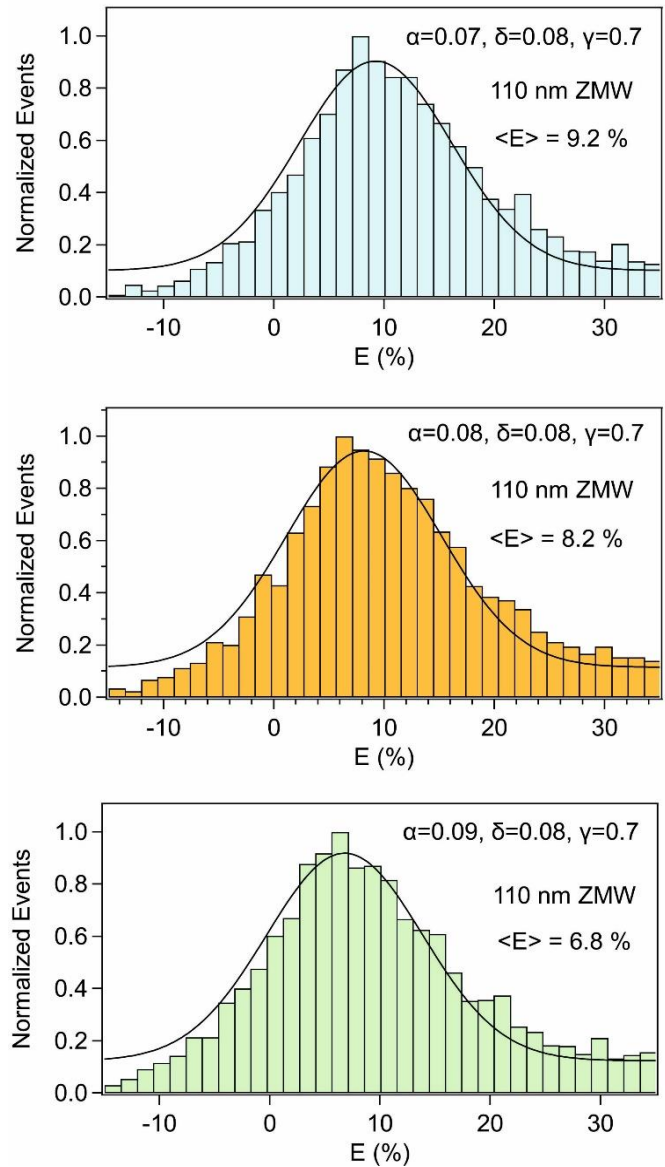

**Figure S5**: Effect of a variation in the crosstalk α parameter on the FRET efficiency histograms of the DNA sample with D-A separation of 13.6 nm in the presence of a 110 nm ZMW. Changing α by ±0.01 modifies the average FRET efficiency by approximately 1%, yet even in the worst case scenario the FRET efficiency found in the ZMW remains 2.5x greater than the one found for the confocal reference. The black line is a Gaussian fit.

**S6. Influence of the direct excitation parameter δ on the measured FRET efficiency**

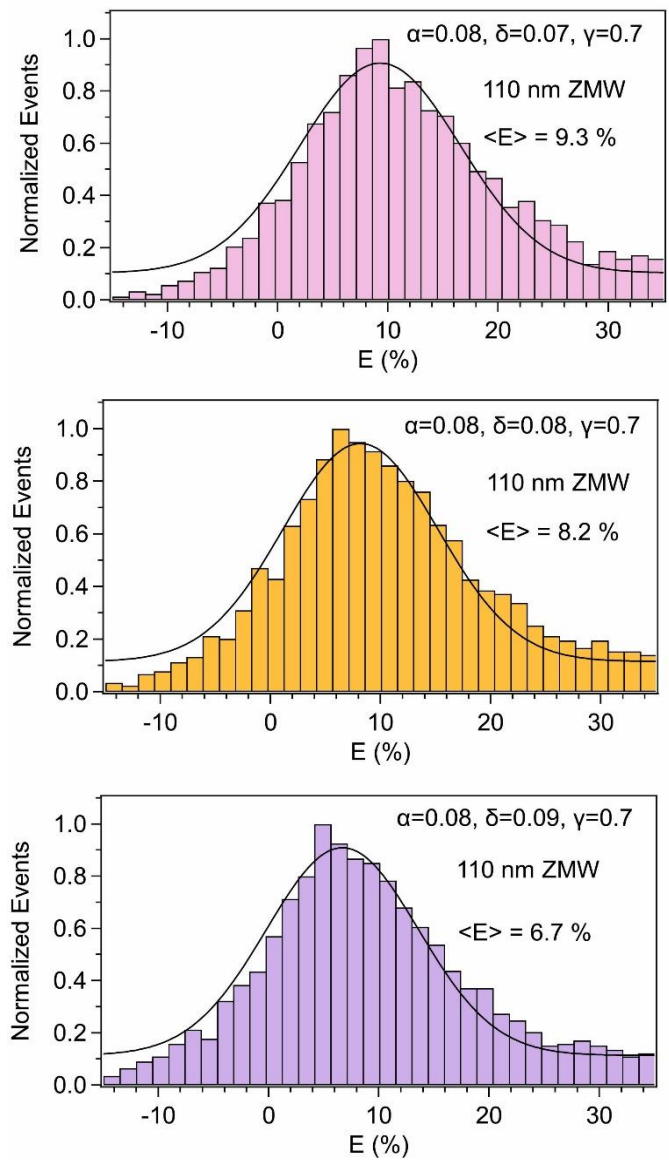

**Figure S6**: Effect of a variation in the direct excitation δ parameter on the FRET efficiency histograms of the DNA sample with D-A separation of 13.6 nm in the presence of a 110 nm ZMW. As for the crosstalk parameter α, we find that changing δ by ±0.01 modifies the average FRET efficiency by approximately 1%. Again, the FRET efficiency in the ZMW remains always 2.5x greater than the one found for the confocal reference. The black line is a Gaussian fit.

**S7. Influence of the γ parameter on the measured FRET efficiency**

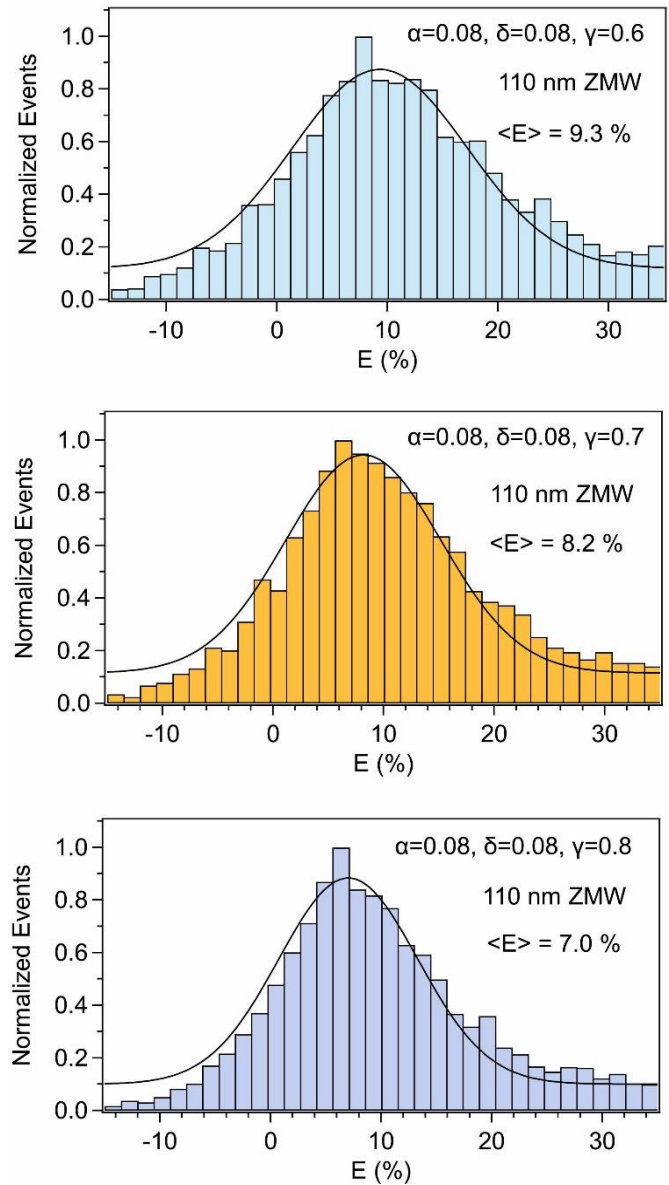

**Figure S7**: Effect of a variation in the γ correction parameter on the FRET efficiency histograms of the DNA sample with D-A separation of 13.6 nm in the presence of a 110 nm ZMW. Changing γ by ±0.1 modifies the average FRET efficiency by approximately 1%. Again, the FRET efficiency in the ZMW remains always 2.5x greater than the one found for the confocal reference. The black line is a Gaussian fit.

## S8. Influence of the binning time on the measured FRET efficiency

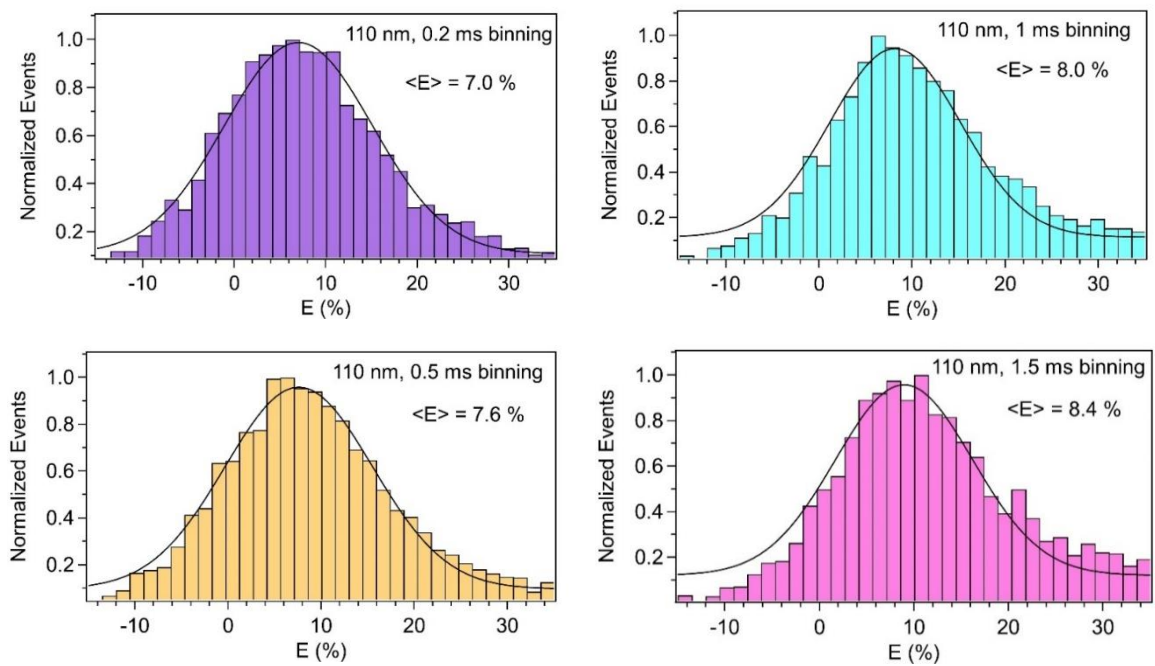

**Figure S8**: Effect of a variation in the binning time used to analyze the PIE-FRET time traces on the FRET efficiency histograms of the DNA sample with D-A separation of 13.6 nm in the presence of a 110 nm ZMW. Using longer binning times tends to increase slightly the average FRET efficiency but without changing our scientific claim. Even for 200 μs binning time, the FRET efficiency in the ZMW is 2.6x larger than the one found for the confocal reference. The black line is a Gaussian fit.

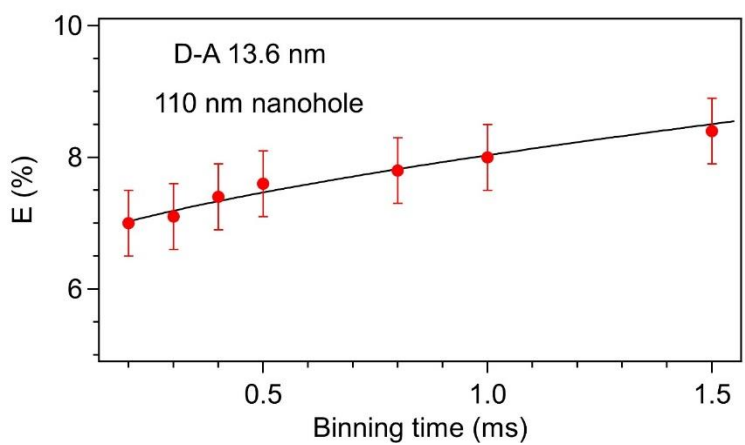

**Figure S9**: Plot of the FRET efficiency deduced from Gaussian fit as a function of the binning time used for the fluorescence time trace analysis. The DNA sample corresponds to a D-A separation of 13.6 nm and the ZMW has a 110 nm diameter.

## S9. Comparison of models to fit the smFRET histograms

Figure S10 compares two different approaches to fit the smFRET histogram. We select here the case of a 100 nm ZMW and 13.6 nm D-A distance (40 base pairs), which goes along with the series displayed in Fig. 3a of the main document. While the Gaussian function provides a simple approach, it tends to overestimate the influence of FRET events below the average value and comparatively underestimate the high FRET events. The data interpolation appears better for the gamma distribution, with flatter residuals and a lower $\chi^2$ value.

With this set of data, one can also compute the statistical average value using the standard probability definition $\sum_i p_i x_i / \sum_i p_i$. For the dataset on Fig. S10, we get a statistical average of 8.5%, in good agreement with the gamma distribution result. As a consequence of the tendency of the Gaussian fit to underestimate the weight of the higher FRET values, the average FRET efficiency deduced from the Gaussian fit is a bit lower (7.1%), yet as this is a rather systematic bias, it affects also the confocal reference and the relative FRET enhancement (ZMW compared to confocal) remains largely unchanged. It is also worth to mention that contrarily to the Gaussian distribution, the average value for the gamma distribution does not occur at the maximum of the distribution but is slightly right-shifted. This may look strange to some readers, but it is simply a consequence of the asymmetric shape of the gamma distribution.

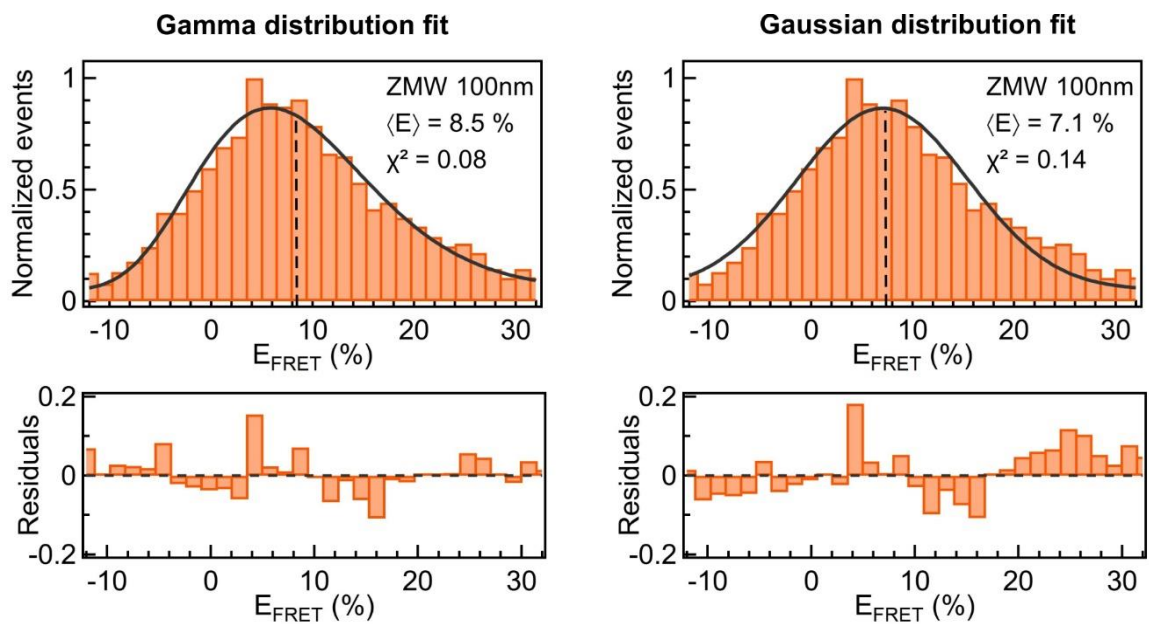

**Figure S10**: Comparison of gamma and Gaussian distributions to fit the smFRET histograms for a 100 nm diameter ZMW and 40 bp D-A separation. The average FRET efficiencies and chi square values are given in each case, the bottom graph represents the residuals from the fits.

## S10. Fluorescence lifetime analysis

The time correlated single photon counting (TCSPC) histograms are fitted by using a Levenberg-Marquard optimization, implemented on a commercial SymPhoTime 64 (PicoQuant GmbH). The model performs an iterative reconvolution fit taking into account the instrument response function (IRF). The time gates in the TCSPC histograms are set to ensure that there are always 92-96% detected photons in the region of interests while fitting the data. The donor fluorescence decay is fitted with a single exponential function for the confocal reference. However, in the case of Al ZMW we found that a biexponential function provides a better fit to the intensity decay. The second lifetime component if fixed at 0.3 ns while fitting the data for the Al ZMW. The average FRET efficiency is determined from the lifetime data by using the equation $E_{FRET} = 1 - \tau_{DA}/\tau_D$, where $\tau_{DA}$ and $\tau_D$ are the fluorescence lifetime of the donor in presence and absence of acceptor. Note that in the case of ZMWs, we use the intensity-averaged lifetime value to determine $E_{FRET}$.

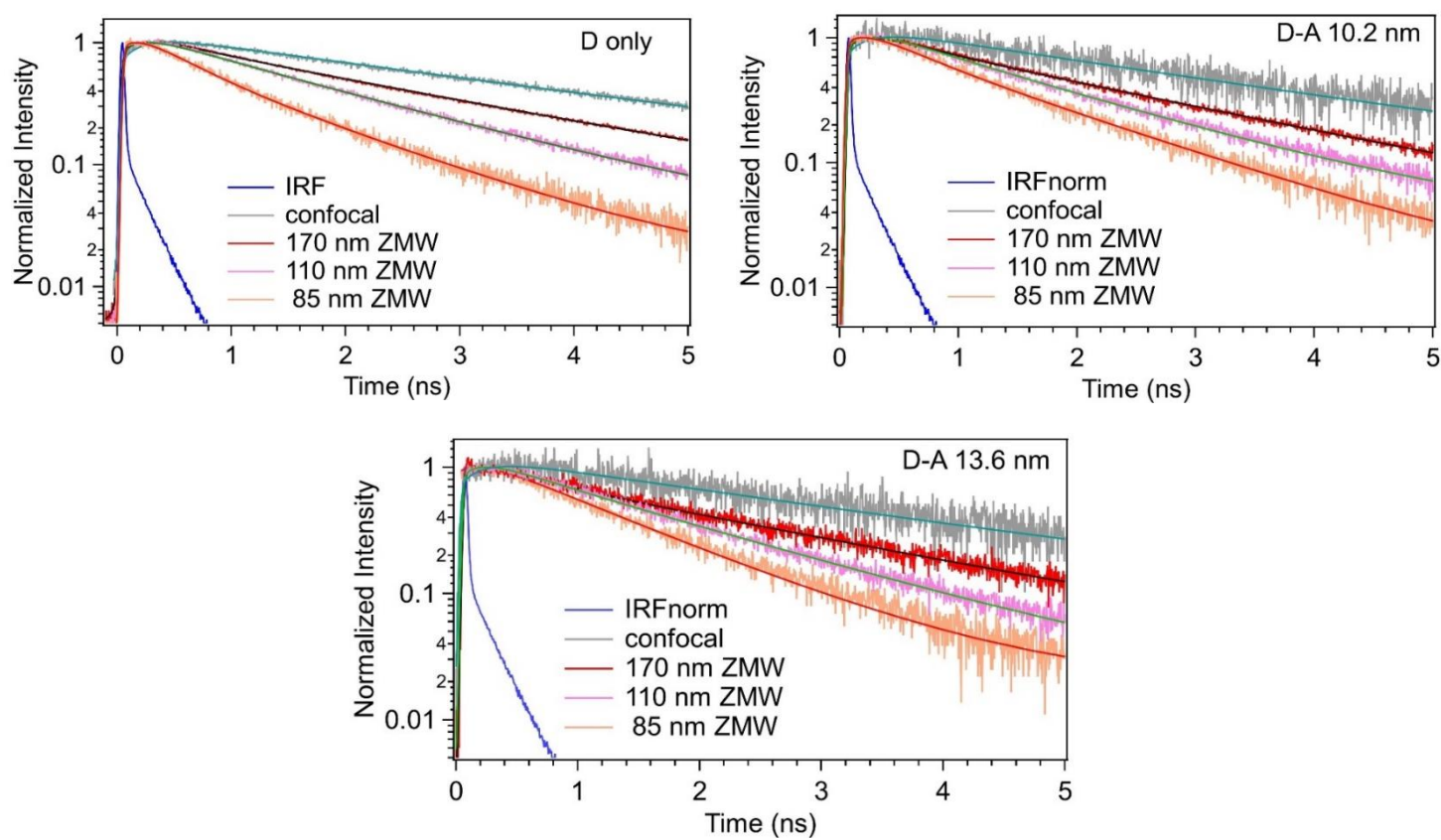

**Figure S11**: Normalized fluorescence intensity decay traces for the Atto 550 donor for different ZMW diameters and the confocal reference. The different datasets correspond to the DNA sample labelled only with the donor dye (D only) and with the acceptor set at different distances (D-A 10.2 nm and D-A 13.6 nm). The thick solid lines are numerical fits. IRF denotes the instrument response function.

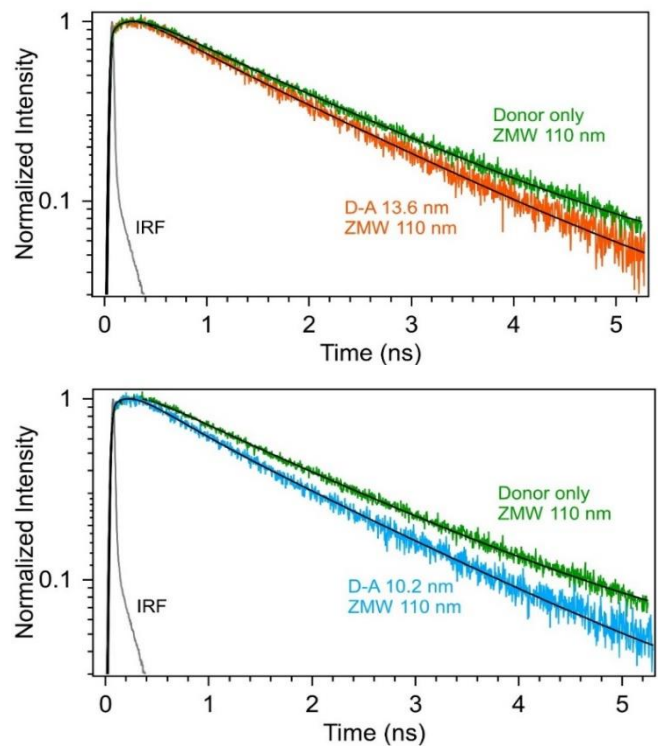

**Figure S12**: Comparison of the donor fluorescence decay traces inside a 110 nm ZMW for the sample labelled only with the donor or the FRET sample where the donor-acceptor separation is 13.6 nm (top panel) and 10.2 nm (bottom panel) respectively. A reduction of the donor decay time in presence of the acceptor confirms the occurrence of the FRET phenomenon. Black lines are numerical fits.

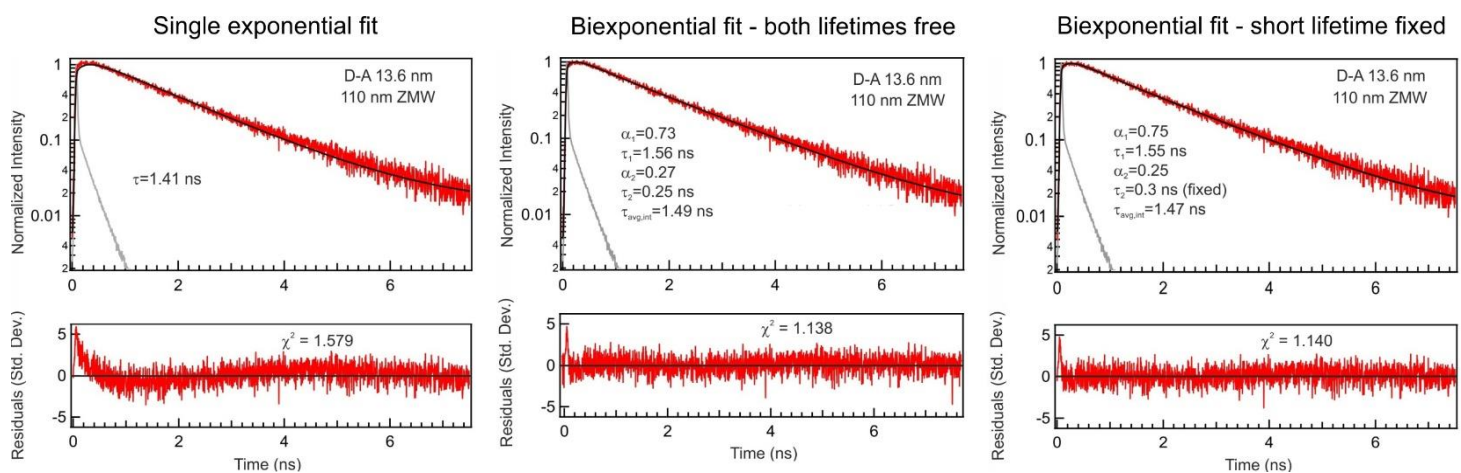

**Figure S13**: Comparison of the fitting procedures to analyze the TCSPC decays. Each panel corresponds to the same dataset (110 nm ZMW, 13.6 nm FRET sample) with a different approach. The bottom traces show the residuals of the fit together with the chi square value.

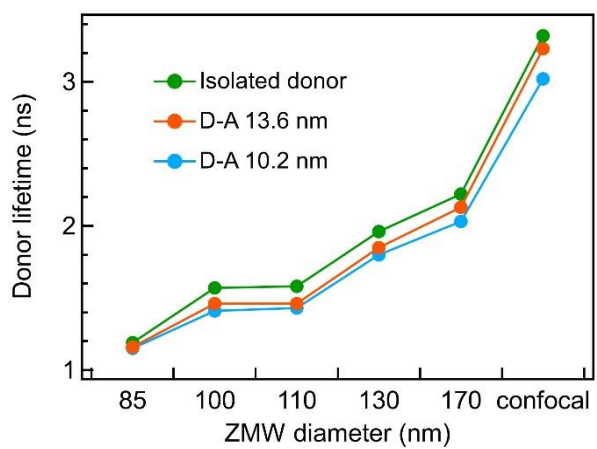

**Figure S14**: Evolution of the intensity-averaged lifetime as a function of the ZMW diameter for the donor-only and the FRET samples. A smaller ZMW diameter reduces the donor lifetime which is further reduced by the presence of the acceptor due to FRET.

### *Fluorescence lifetime data for the D-A 13.6 nm dsDNA*

**Table S1**: Results obtained from the biexponential fit to the TCSPC histograms for the DNA sample labelled only with the donor in presence of ZMWs of different diameters. 92%-96% photons were considered for the fit. The second lifetime component is fixed at 0.3 ns.

| Diameter / nm | $\tau_1$/ns | $\tau_2$/ns | $A_1$ | $A_2$ | $<\tau_D>$ / ns | $\chi^2$ |
|---|---|---|---|---|---|---|
| 170 | 2.28 | 0.3 | 0.94 | 0.23 | 2.22 | 1.112 |
| 130 | 2.00 | 0.3 | 1.82 | 0.30 | 1.96 | 0.956 |
| 110 | 1.63 | 0.3 | 2.54 | 0.52 | 1.58 | 1.071 |
| 100 | 1.64 | 0.3 | 3.07 | 0.90 | 1.57 | 1.064 |
| 85 | 1.30 | 0.3 | 0.19 | 0.11 | 1.19 | 1.305 |

**Table S2**: Results of the biexponential fit to the TCSPC histograms for 13.6 nm FRET sample in presence of ZMWs of different diameters. 92%-96% photons were considered for the fit. The second lifetime component is fixed at 0.3 ns.

| Diameter / nm | $\tau_1$/ns | $\tau_2$/ns | $A_1$ | $A_2$ | $<\tau_{DA}>$ / ns | $\chi^2$ |
|---|---|---|---|---|---|---|
| 170 | 2.29 | 0.3 | 0.36 | 0.24 | 2.13 | 1.181 |
| 130 | 1.92 | 0.3 | 1.32 | 0.35 | 1.85 | 1.271 |
| 110 | 1.54 | 0.3 | 0.79 | 0.27 | 1.46 | 1.208 |
| 100 | 1.56 | 0.3 | 0.43 | 0.19 | 1.46 | 1.433 |
| 85 | 1.26 | 0.3 | 0.45 | 0.21 | 1.16 | 1.099 |

**Table S3:** Average FRET efficiency $E^{lifetime}$ for the DA 13.6 nm DNA sample determined from the fluorescence lifetime of the donor in absence ($\tau_D$) and in presence of acceptor ($\tau_{DA}$) by using the formula $E = 1 - \frac{\tau_{DA}}{\tau_D}$. The lifetimes are the intensity-averaged values determined in Table S1 and S2. The relative error associated with each determined value is estimated to be around 10%. These values are found in excellent agreement with the ones determined with the PIE-FRET burst analysis, which further confirms the validity of our conclusions (see Fig. 3c for a graphical display).

| Diameter / nm | $\tau_{DA}$/ns* | $\tau_D$/ns* | $E^{lifetime}$ (%) | $E^{PIE-FRET}$ (%) |
|---|---|---|---|---|
| confocal | 3.23 | 3.32 | 2.7 | 2.7 |
| 170 | 2.13 | 2.22 | 4.1 | 3.6 |
| 130 | 1.85 | 1.96 | 5.6 | 6.0 |
| 110 | 1.46 | 1.58 | 7.6 | 8.0 |
| 100 | 1.46 | 1.57 | 7.0 | 6.9 |
| 85 | 1.16 | 1.19 | 2.5 | 2.2 |

*Fluorescence lifetime data for the D-A 10.2 nm dsDNA*

**Table S4**: Results of the biexponential fit to the TCSPC histograms for 10.2 nm FRET sample in presence of ZMWs of different diameters. 92%-96% photons were considered for the fit. The second lifetime component is fixed at 0.3 ns.

| Diameter / nm | $\tau_1$/ns | $\tau_2$/ns | $A_1$ | $A_2$ | $<\tau_{DA}>$ / ns | $\chi^2$ |
|---|---|---|---|---|---|---|
| 170 | 2.14 | 0.3 | 2.62 | 1.15 | 2.03 | 1.100 |
| 130 | 1.92 | 0.3 | 1.24 | 0.59 | 1.80 | 1.041 |
| 110 | 1.49 | 0.3 | 1.47 | 0.38 | 1.43 | 1.142 |
| 100 | 1.52 | 0.3 | 0.93 | 0.46 | 1.41 | 0.962 |
| 85 | 1.20 | 0.3 | 1.06 | 0.328 | 1.15 | 1.267 |

**Table S5**: Average FRET efficiency $E^{lifetime}$ for the DA 10.2 nm DNA sample determined from the fluorescence lifetime of the donor in absence ($\tau_D$) and in presence of acceptor ($\tau_{DA}$). The lifetimes are the intensity-averaged values determined in Table S1 and S4. The relative error associated with each determined value is estimated to be around 10%. Again, these values stand in excellent agreement with the results from the PIE-FRET burst analysis (see Fig. 3d for a graphical display).

| Diameter / nm | $\tau_{DA}$/ns* | $\tau_D$/ns* | $E^{lifetime}$ (%) | $E^{PIE-FRET}$ (%) |
|---|---|---|---|---|
| confocal | 3.02 | 3.32 | 9.0 | 8.9 |
| 170 | 2.03 | 2.22 | 8.6 | 8.6 |
| 130 | 1.80 | 1.96 | 8.2 | 7.8 |
| 110 | 1.43 | 1.58 | 9.5 | 10.1 |
| 100 | 1.41 | 1.57 | 10.2 | 9.5 |
| 85 | 1.15 | 1.19 | 3.3 | 3.9 |

**S11. S-E plot diagrams**

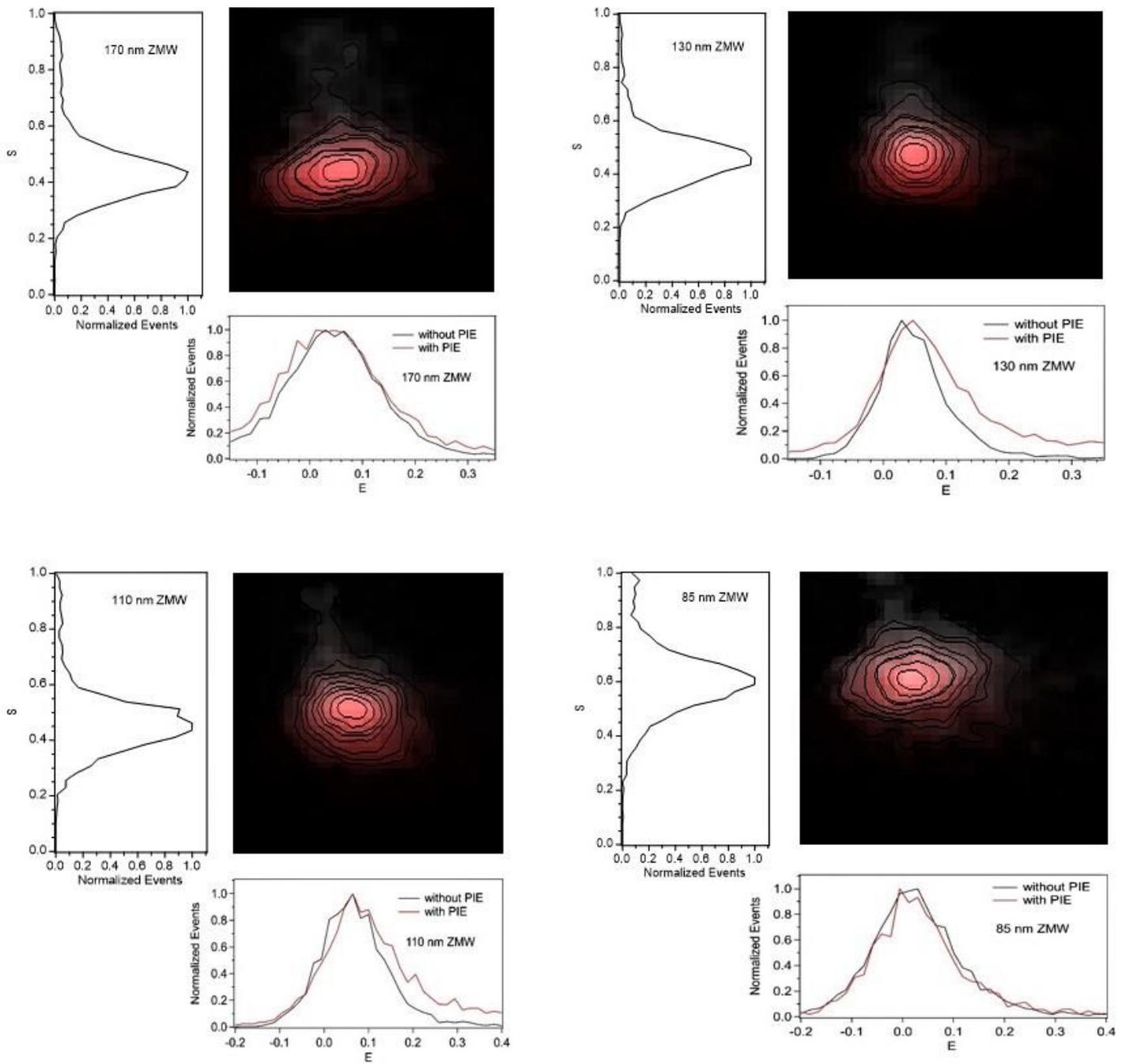

**Figure S15**: S-E plots for the DNA sample with 13.6 D-A separation inside ZMWs of different diameters.

## S12. Experimental determination of the $\gamma$ factor

The $\gamma$ correction factor can be determined experimentally using the measurements of the photon stoichiometry S and the average FRET efficiency. The stoichiometry S is defined as

$$S = \frac{(n_A^{green} - \alpha\, n_D^{green} - \delta\, n_A^{red}) + \gamma\, n_D^{green}}{(n_A^{green} - \alpha\, n_D^{green} - \delta\, n_A^{red}) + \gamma\, n_D^{green} + n_A^{red}} \tag{S2}$$

where $n_D^{green}$ and $n_A^{green}$ are the numbers of photons per burst detected in the donor and acceptor channels following a green excitation pulse, and $n_A^{red}$ is the number of photons per burst detected in the acceptor channel following a red excitation pulse. Equation (S2) can be further simplified to introduce the FRET efficiency (as defined by Eq. (4) of the main document):

$$\frac{1}{S} = 1 + (1 - E_{FRET})\, \frac{1}{\gamma}\, \frac{n_A^{red}}{n_D^{green}} \tag{S3}$$

Equation (S3) can be simplified even further by reminding that by definition of the FRET efficiency, $n_D^{green} = (1 - E_{FRET})\, n_{D,donor\ only}^{green}$, where $n_{D,donor\ only}^{green}$ is the number of photons per burst detected in the donor channel following a green excitation pulse for the sample labelled only with the donor.

To improve the accuracy of the analysis, we replace the number of photons per bursts $n_A^{red}$ and $n_{D,donor\ only}^{green}$ by the average brightness $CRM_A^{red}$ and $CRM_{D,donor\ only}^{green}$ measured by FCS over the full trace duration (CRM stands for count rate per molecule and is obtained in FCS by the ratio of the average fluorescence intensity by the average number of molecules detected). Equation (S3) can then be simplified to express the $\gamma$ correction factor as function of experimentally determined values:

$$\gamma = \frac{S}{1-S}\, \frac{CRM_A^{red}}{CRM_{D,donor\ only}^{green}} \tag{S4}$$

Our results are summarized in Table S6 and Fig. S16. The $\gamma$ values determined from the stoichiometry S and the FCS brightness are in excellent agreement with the values deduced using the approach in Eq. (5) of the main document. This holds for both the confocal configuration and the different ZMWs.

**Table S6:** Experimental determination of the $\gamma$ factor for the DNA sample with D-A 13.6 nm using the approach in Eq. (S4) to obtain the $\gamma_{stoichiometry}$ values or the approach in Eq. (5) to obtain $\gamma_{quantum\ yield}$.

| Diameter (nm) | S | $\gamma_{stoichiometry}$ | $\gamma_{quantum\ yield}$ |
|---|---|---|---|
| confocal | 0.55 | 0.71 | 0.80 |
| 170 | 0.42 | 0.67 | 0.71 |
| 130 | 0.45 | 0.84 | 0.80 |
| 110 | 0.47 | 0.77 | 0.70 |
| 100 | 0.45 | 0.77 | 0.70 |
| 85 | 0.60 | 0.81 | 0.80 |

**Table S7:** Count rate per molecule for the donor only (DO) and acceptor only (AO) DNA sample in presence of ZMWs of different diameters.

| Diameter (nm) | $CRM_{DO}$ (kHz) | $CRM_{AO}$ (kHz) |
|---|---|---|
| confocal | 15.9 | 9.2 |
| 170 | 65.9 | 61.5 |
| 130 | 81.4 | 83.6 |
| 110 | 134.2 | 116.2 |
| 100 | 120.1 | 113.1 |
| 85 | 81.6 | 66.0 |

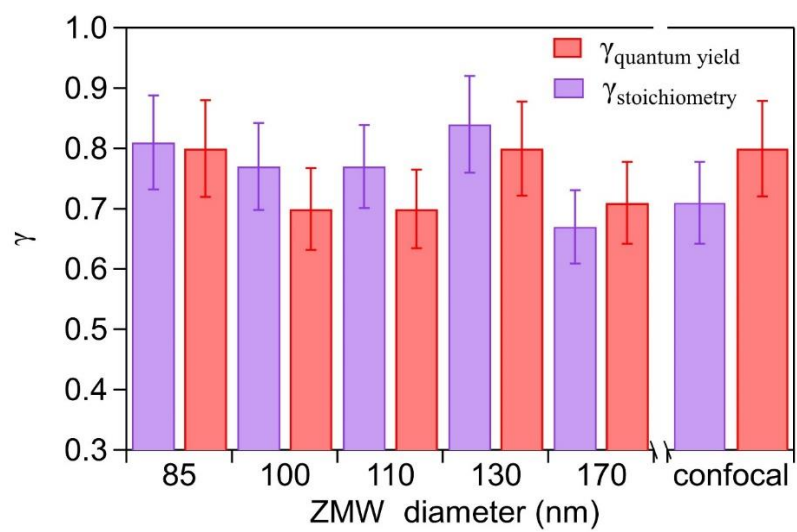

**Figure S16**: Plot of γ value determined from stoichiometry analysis (Eq. S4) and from the quantum yield and detection sensitivities of the Atto 550 and Atto 647N fluorophore as a function of ZMW diameter (Eq. 5). Both γ estimates are consistent with each other within the experimental uncertainties.

**S13. Fluorescence spectra of the multi-acceptor DNA sample**

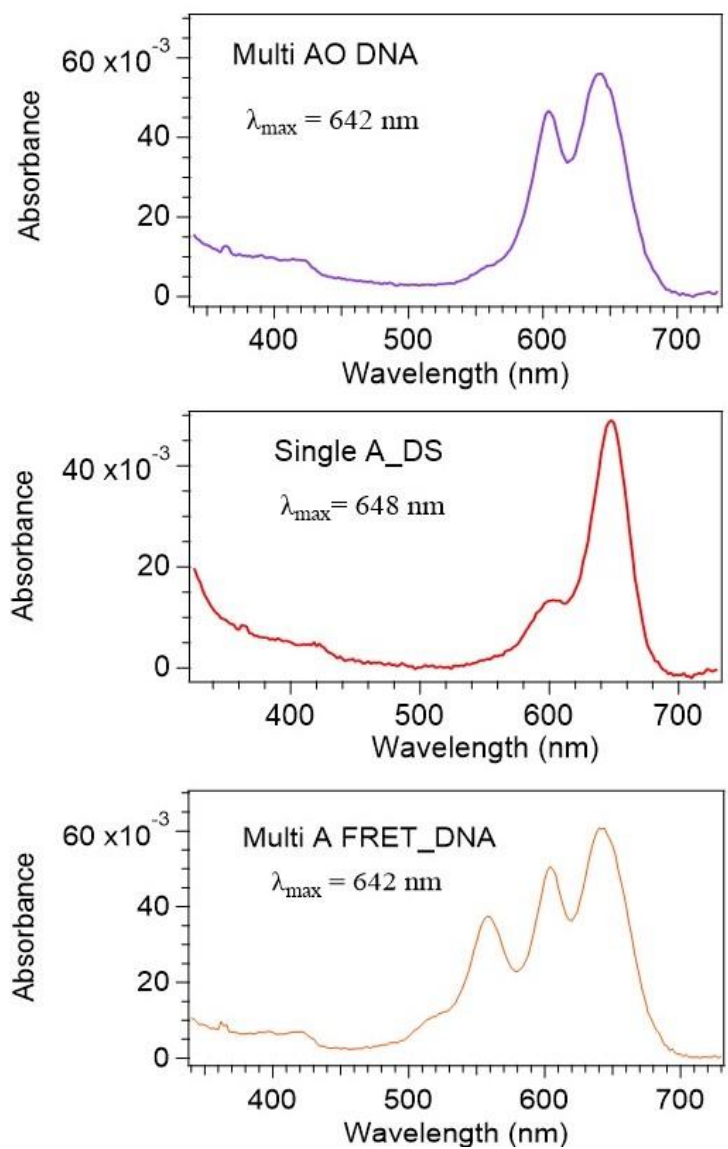

**Figure S17**: Absorption spectra of the multi acceptor without donor (multi AO), single acceptor without donor (single AO) and multi acceptor FRET sample. The absorption of the peak at 604 nm in higher for multi AO and multi A FRET as compared to single acceptor indicating ground state complex formation. The multi A FRET sample also shows an absorption band at 558 nm due to the presence Atto 550 donor.

*Fluorescence emission spectra*

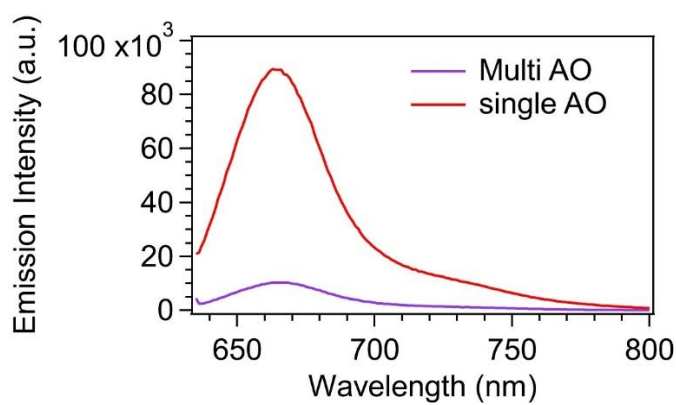

**Figure S18**: Emission spectra of the multi acceptor without donor (multi AO) and single acceptor without donor (single AO). The excitation wavelength is 600 nm. No change in the shape of the emission spectrum is observed for both DNA sample. However, the emission is heavily quenched in case of multi acceptors. The concentration for both DNA samples is equivalent.

*Fluorescence excitation spectra*

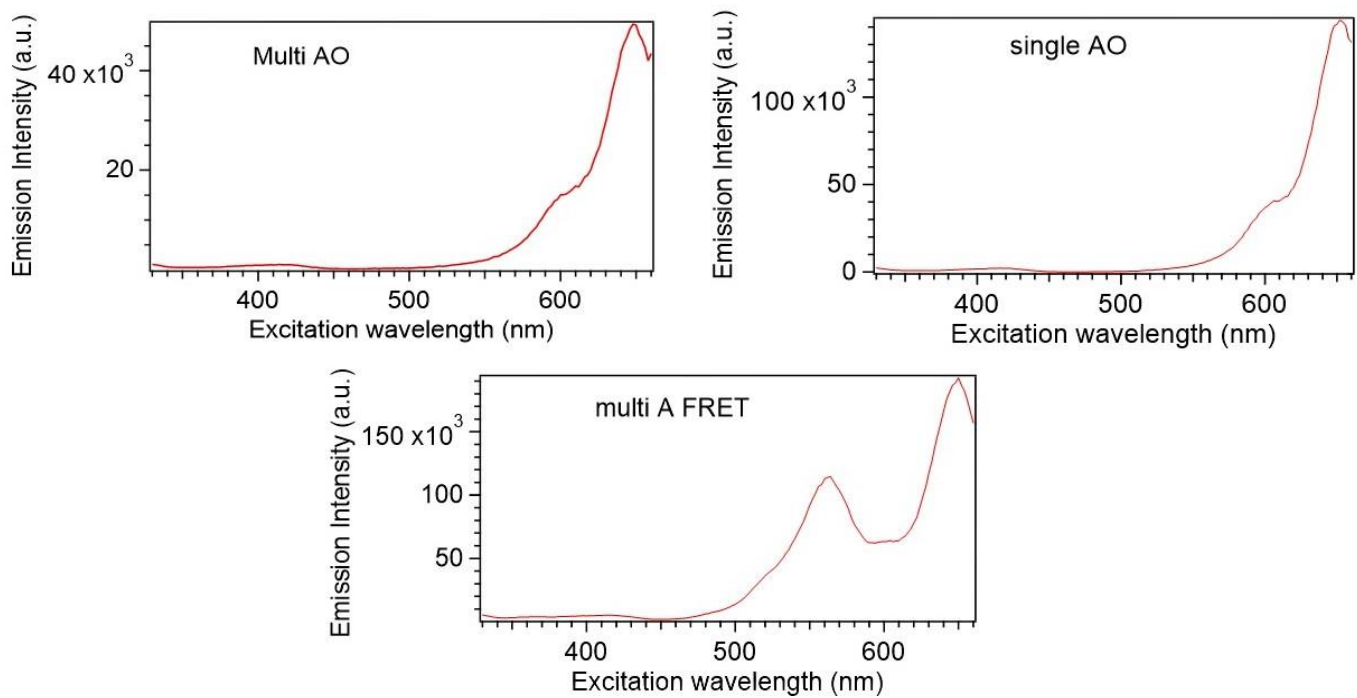

**Figure S19:** Excitation spectra of the multi acceptor without donor (multi AO), single acceptor without donor (single AO) and multi acceptor FRET sample. The excitation spectra of multi AO and single AO have identical shapes (but different intensities) that confirming similar absorbing species contribute to the emission. In case of the FRET sample, the excitation spectrum reveals that both donor and acceptor absorption give rise to the acceptor emission.

## S14. Expression of the total energy transfer rate constant inside the ZMW

In this section, we discuss how we derive the Eq. (2) of the main text as $\Gamma_{\text{FRET}}^{\text{tot}} = \Gamma_{\text{FRET}}^{0} + \Gamma_{\text{FRET}}^{\text{ZMW}}$. The starting point is to remind that the total FRET rate constant inside the ZMW $\Gamma_{\text{FRET}}^{\text{tot}}$ is proportional to the power transferred from the donor to the acceptor. [S4, S5] Within the semi-classical general description of dipole-dipole interaction, the power transferred by a donor D to a polarizable acceptor A can be written as [S4, S6]

$$P_{D \to A} = \frac{\omega}{2} \, Im\{\alpha_A\} \left| \boldsymbol{n_A} \cdot \boldsymbol{E_{D,tot}(r_A)} \right|^2, \tag{S5}$$

where $\overrightarrow{\alpha_A} = \alpha_A \, \boldsymbol{n_A} \boldsymbol{n_A}$ is the acceptor's polarizability tensor and $\boldsymbol{E_{D,tot}(r_A)}$ is the electric field due to the donor dipole at the acceptor position. Hence the total FRET rate constant inside the ZMW $\Gamma_{\text{FRET}}^{\text{tot}}$ is proportional to the modulus square of the electric field emitted by the donor dipole at the acceptor dipole position.

The next step is to write the electric field emitted by the donor dipole as a sum of the reference field in the absence of structure $\boldsymbol{E_{D,ref}(r_A)}$ plus a contribution from the field scattered by the ZMW structure $\boldsymbol{E_{D,sca}(r_A)}$ :

$$\boldsymbol{E_{D,tot}(r_A)} = \boldsymbol{E_{D,ref}(r_A)} + \boldsymbol{E_{D,sca}(r_A)} \tag{S6}$$

This is a widely used approach in nanophotonics while computing the LDOS, [S4] and it remains always valid as it basically serves to define the scattered field $\boldsymbol{E_{D,sca}(r_A)}$. The major difference here with the LDOS calculation is that the electric field is considered not at the position of the donor source but at the position of the acceptor's position $\boldsymbol{r_A}$.

Combining Eqs. (S5) and (S6), and not writing the obvious position dependence $\boldsymbol{r_A}$, the total FRET rate can be expanded into a sum of terms:

$$\begin{aligned} \Gamma_{\text{FRET}}^{\text{tot}} = K \left| \boldsymbol{n_A} \cdot \left( \boldsymbol{E_{D,ref}} + \boldsymbol{E_{D,sca}} \right) \right|^2 = \; & K \left| \boldsymbol{n_A} \cdot \boldsymbol{E_{D,ref}} \right|^2 + K \left| \boldsymbol{n_A} \cdot \boldsymbol{E_{D,sca}} \right|^2 \\ & + 2 \, K \, Re\left( \boldsymbol{n_A} \cdot \boldsymbol{E_{D,ref}} \; \boldsymbol{n_A} \cdot \boldsymbol{E_{D,sca}}^* \right) \end{aligned} \tag{S7}$$

Here $K$ is a just proportionality factor to shorten the equation. As pointed out in Ref. [S7], Equation (S7) shows that the FRET rate constant inside any nanophotonic structure can be actually seen as an interference pattern between the reference field radiated in an homogeneous medium $\boldsymbol{E_{D,ref}(r_A)}$ and the field scattered by the nanostructure $\boldsymbol{E_{D,sca}(r_A)}$.

Importantly, the first term in the right side of Eq. (S7) corresponds the FRET rate constant in the absence of the structure $\Gamma_{\text{FRET}}^{0} = K \left| \boldsymbol{n_A} \cdot \boldsymbol{E_{D,ref}} \right|^2$. This contribution is the one following the $1/R^6$ distance dependence. Then we can define the remaining terms in the right side of Eq. (S7) as the contribution to the FRET rate stemming from the nanostructure:

$$\Gamma_{\text{FRET}}^{\text{ZMW}} = K \left| \boldsymbol{n_A} \cdot \boldsymbol{E_{D,sca}} \right|^2 + 2 \, K \, Re\left( \boldsymbol{n_A} \cdot \boldsymbol{E_{D,ref}} \; \boldsymbol{n_A} \cdot \boldsymbol{E_{D,sca}}^* \right) \tag{S8}$$

So that we retrieve the additive contribution $\Gamma_{\text{FRET}}^{\text{tot}} = \Gamma_{\text{FRET}}^{0} + \Gamma_{\text{FRET}}^{\text{ZMW}}$ used to simplify the expressions in the main text. It should be pointed out here that $\Gamma_{\text{FRET}}^{\text{ZMW}}$ is not just a constant term, but it depends on the interference between the reference and the scattered field. Even if $\boldsymbol{E_{D,sca}}$ can be quite small, the interference cross term with $\boldsymbol{E_{D,ref}}$ may become important.

**Supplementary references**